\documentclass[12pt,preprint]{aastex}
\usepackage{graphicx}

\shorttitle{ATCA and $Spitzer$ observations of CG\,30 and BHR\,71}
\shortauthors{Chen et al.}

\begin{document}

\title{ATCA and \emph{Spitzer} Observations of the Binary Protostellar Systems CG\,30 and
BHR\,71}

\author{Xuepeng Chen, Ralf Launhardt}
\affil{Max Planck Institute for Astronomy, K\"{o}nigstuhl 17,
D-69117 Heidelberg, Germany; chen@mpia.de}

\and

\author{Tyler L. Bourke}
\affil{Harvard-Smithsonian Center for Astrophysics, 60 Garden
Street, Cambridge, MA 02138, USA}

\and

\author{Thomas Henning}
\affil{Max Planck Institute for Astronomy, K\"{o}nigstuhl 17,
D-69117 Heidelberg, Germany}

\and

\author{Peter J. Barnes}
\affil{School of Physics A28, University of Sydney, Sydney
NSW2006, Australia}

\begin{abstract}

We present interferometric observations with resolution of
$\sim$\,3$''$ of the isolated, low-mass protostellar double cores
CG\,30 and BHR\,71 in the N$_{2}$H$^{+}$\,(1\,$-$\,0) line and at
3\,mm dust continuum, using the Australian Telescope Compact Array
(ATCA). The results are complemented by infrared data from the
{\it Spitzer Space Telescope}. In CG\,30, the 3\,mm dust continuum
images resolve two compact sources with a separation of
$\sim$\,21\farcs7 ($\sim$\,8700\,AU). In BHR\,71, strong dust
continuum emission is detected at the position of the mid-infrared
source IRS1, while only weak emission is detected from the
secondary mid-infrared source IRS2. Assuming standard gas to dust
ratio and optically thin 3\,mm dust continuum emission, we derive
hydrogen gas masses of 0.05\,$-$\,2.1\,$M_\odot$ for the four
sub-cores. N$_{2}$H$^{+}$\,(1\,$-$\,0) line emission is detected
in both CG\,30 and BHR\,71, and is spatially associated with the
thermal dust continuum emission. By simultaneously fitting the
seven hyperfine line components of N$_{2}$H$^{+}$, we derive the
velocity fields and find symmetric velocity gradients in both
sources. Assuming that these gradients are due to core rotation,
we estimate the specific angular momenta and ratios of rotational
energy to gravitational energy for all cores. Estimated virial
masses of the sub-cores range from 0.1\,$-$\,0.6\,$M_\odot$. We
also find that the N$_{2}$H$^{+}$ emission is strongly affected by
the outflows, both in terms of entrainment and molecule
destruction. $Spitzer$ images show the mid-infrared emission from
all four sub-cores, which is spatially associated with the 3\,mm
dust continuum emission. All four sources appear to drive their
own outflows, as seen in the shock-excited 4.5\,$\mu$m images.
Based on the ATCA and $Spitzer$ observations, we construct
spectral energy distributions (SEDs) and derive temperatures and
luminosities for all cores. The analysis of the SEDs suggests an
evolutionary discrepancy between the two sub-cores in both CG\,30
and BHR\,71, which could be due to effects of relative inclinations. 
Based on the morphology and velocity
structure, we suggest that the sub-cores in CG\,30 were formed by
initial fragmentation of a filamentary prestellar core, while
those in BHR\,71 could originate from rotational fragmentation of
a single collapsing protostellar core.

\end{abstract}


\section{INTRODUCTION}

Although statistical properties of binary stars have been
determined over the past two decades (see e.g., Reipurth et al.
2007 for recent reviews), many key questions concerning their
origin are still poorly understood. What is the formation
mechanism of binary/multiple systems? How are mass and angular
momentum distributed during their formation? What is the
difference between cores forming binaries and those forming single
stars? To answer these questions, direct observations of the
earliest, deeply embedded phase of binary star formation are
needed. This phase is unfortunately not accessible to optical and
near-infrared (NIR) wavelengths, due to the large amounts of
circumstellar material present. Observations of the gas and
optically thin dust emission at millimeter (mm) wavelengths are
therefore needed, to probe the system kinematics and individual
envelope masses. However, these observations were long hampered by
the low angular resolution of mm telescopes, and only recently
have the earliest phases of binary star formation been
observationally identified and studied in detail thanks to the
availability of large (sub-) mm interferometers, although the
number of known systems is still very small (Looney et al. 2000;
Launhardt 2004).

To search for binary protostars and to derive their kinematic
properties, we have started a systematic program to observe, at
high angular resolution, a number of isolated low-mass prestellar
and protostellar molecular cloud cores. The initial survey was
conducted at the Owens Valley Radio Observatory (OVRO) mm array
(Launhardt 2004; Chen, Launhardt, \& Henning 2007, hereafter
Paper\,I; Launhardt et al., in prep.), and is now continued with
the Australia Telescope Compact Array (ATCA) and the IRAM Plateau
de Bure Interferometer (PdBI) array. In this paper we present ATCA
observations of two southern protobinaries in the
N$_{2}$H$^{+}$\,(1\,$-$\,0) molecular line and at 3\,mm dust
continuum, together with complementary mid-infrared (MIR) data
from the \emph{Spitzer Space Telescope} (hereafter $Spitzer$).

CG\,30 (also known as BHR\,12 or DC 253.3$-$1.6) is a
bright-rimmed cometary globule located in the Gum Nebula region.
The distance towards CG\,30 is somewhat uncertain, with estimates
ranging from 200\,pc (Knude et al. 1999) to 400\,pc (Brandt 1971;
Reipurth 1983). For consistency with earlier papers (e.g., Henning
\& Launhardt 1998), we use here 400\,pc. The globule harbors an
elongated protostellar core as seen in the single-dish mm dust
continuum image (Launhardt et al. in prep.). Higher resolution
submm continuum observations (SCUBA) resolve the source into two
sub-cores with a projected separation of $\sim$\,20$''$
($\sim$\,8000\,AU) and masses of 0.17\,$\pm$\,0.05 and
0.14\,$\pm$\,0.05 $M_\odot$ (Henning et al. 2001). The northern
core is associated with a NIR source, which drives the Herbig-Haro
flow HH\,120 (see Hodapp \& Ladd 1995 and references therein). The
newly discovered southern core is the origin of a protostellar jet
with position angle (P.A.) 44$^\circ$ (Hodapp \& Ladd 1995), but
no NIR source is seen at this position (see Launhardt et al.
2001).

BHR\,71 (also known as DC 297.7$-$2.8) is an isolated Bok globule
located at a distance of $\sim$ 200\,pc (Bourke et al. 1997;
hereafter B97). A highly-collimated bipolar outflow, which is
lying almost in the plane of the sky, was discovered by CO
observations in this region. The driving source is associated with
IRAS\,11590$-$6452 and was classified as a Class\,0 protostar with
a total luminosity of $\sim$ 9\,$L_\odot$ (B97). ISOCAM
observations have revealed that the IRAS source is associated with
two embedded protostars, IRS1 and IRS2, with a projected
separation of $\sim$\,17$''$ ($\sim$\,3400\,AU; Bourke 2001;
hereafter B01). IRS1 and IRS2 each drive a CO outflow: the
well-known large-scale collimated bipolar outflow is driven by
IRS1 and another fainter and smaller bipolar outflow is driven by
IRS2 (see B01 and Parise et al. 2006). Only IRS1 appears to be
associated with a substantial amount of circumstellar material,
but neither is directly detected at NIR wavelengths (B01).

\section{OBSERVATIONS AND DATA REDUCTION}

\subsection{ATCA Observations}

Millimeter interferometric observations at 95\,GHz of CG\,30 and
BHR\,71 were carried out using ATCA with five 22\,m telescopes in
May and August 2005. Observations were obtained in two different
array configurations (H168 and H75) with projected baselines
ranging from 22 to 180\,m. All antennas were equipped with cooled
SIS receivers, which provided average system temperatures of
200\,$-$\,350\,K at the observing frequency. A digital correlator
was used with 2 independent spectral windows. The narrow window
(bandwidth $\sim$\,8\,MHz), with a channel width of 0.019\,MHz,
was centered on the N$_{2}$H$^{+}$\,(1\,$-$\,0) line at
93.17\,GHz\footnote{During the observations the sky frequency
changed by less than one channel due to the lack of doppler
tracking at ATCA, and corrections were applied offline to obtain
correct frequencies and LSR velocities}. The broad window
(bandwidth $\sim$ 128\,MHz) was centered at 95.0\,GHz and was used
to measure the 3.1\,mm dust continuum emission simultaneously with
N$_{2}$H$^{+}$. The two sources were observed with 2-point mosaics
each. The primary beam size at 93\,GHz is $\sim$ 38$''$. Amplitude
and phase were calibrated through frequent observations of quasars
nearby to each source (0745$-$330 for CG\,30 and 1057$-$797 for
BHR\,71), typically every 20 minutes, resulting in an absolute
position uncertainty of $\leq$\,0\farcs2. Flux densities were
calibrated using the secondary calibrator 1253$-$055, the flux of
which was regularly compared to Uranus and adopted as 19.0\,Jy for
May observations (H168 configuration) and 14.7\,Jy for August
observations (H75 configuration). Additional effort was made to
improve the gain-elevation calibration of the antennas, which can
significantly affect the flux density scale, especially when
observing at high elevation. The estimated total flux uncertainty
is $<$\,20\%. Observing parameters are summarized in Table~1.

The data were calibrated and images produced using MIRIAD (Sault
et al. 1995) and its CLEAN algorithm, with ``robust" $uv$
weighting parameter +1 (Briggs et al. 1999). Synthesized beam
sizes are 3$''$\,$-$\,4$''$. Noise levels (1\,$\sigma$ rms) in the
final maps are 0.5\,$-$\,2 mJy/beam for the continuum and
20\,$-$\,65 mJy/beam for the N$_{2}$H$^{+}$ line (see Table~1).
Further analysis and figures were done with the
GILDAS\footnote{http://www.iram.fr/IRAMFR/GILDAS} software
package.

\subsection{\emph{Spitzer} Observations}

Mid-infrared data of CG\,30 and BHR\,71 were obtained from the
$Spitzer$ Science Center\footnote{http://ssc.spitzer.caltech.edu}.
CG\,30 was observed on 2004 April 9 with the Multiband Imaging
Photometer for $Spitzer$ (MIPS: AOR key 9426688) and May 26 with
the Infrared Array Camera (IRAC; AOR key 5097216). BHR\,71 was
observed on 2004 June 10 with IRAC (AOR key 5107200) and 2005
March 7 with MIPS (AOR key 9434112). Both sources were observed as
part of the c2d Legacy program (Evans et al. 2003).

The data were processed by the $Spitzer$ Science Center using
their standard pipeline (version S14.0) to produce Post Basic
Calibrated Data (P-BCD) images, which are flux-calibrated into
physical units (MJy sr$^{-1}$). Flux densities in the IRAC bands
were measured with aperture photometry in the IRAF APPHOT package,
using the radii, background aperture annuli, and aperture
corrections recommended by the $Spitzer$ Science Center. The
results were compared to c2d, which used PSF fitting, and found to
be within the uncertainties. Flux densities in the MIPS bands were
measured with GILDAS because sources in the MIPS images are not
fully resolved (see $\S$\,3.3). Further analysis and figures were
completed with GILDAS.

\section{RESULTS}
\subsection{Dust Continuum}

The 3\,mm dust continuum image of CG\,30 (Fig.\,1a) shows two
compact sources with an angular separation of
21\farcs7\,$\pm$\,0\farcs6, corresponding to a projected linear
separation of 8700\,$\pm$\,240\,AU at a distance of 400\,pc.
Following Henning et al. (2001), we refer to the northern source as
CG\,30N and to the southern source as CG\,30S. From Gaussian $uv$
plane fitting, we derive flux densities of
15.8\,$\pm$\,3.2\,mJy\footnote{The error bar is derived from
$\sqrt{\sigma^2_{\rm cali}+\sigma^2_{\rm fit}}$, where
$\sigma_{\rm cali}$ is the uncertainty from calibration
($\sim$\,20\% of flux density) and $\sigma_{\rm fit}$ is the
uncertainty from Gaussian fitting.} for source N and
6.0\,$\pm$\,1.3\,mJy for source S. The large-scale common
envelope, detected in the submm single-dish maps with a radius of
$\sim$\,14000\,AU and a flux density of $\sim$\,7.4\,Jy (Henning
et al. 2001), is resolved out by the interferometer at 3\,mm.
Source positions and deconvolved FWHM sizes of the two embedded
sources, derived from Gaussian $uv$ plane fitting, are listed in
Table~2.

In the 3\,mm dust continuum image of BHR\,71 (Fig.\,1b), strong
emission is detected at the position of IRS1, and only weak
emission ($\sim$\,3\,$\sigma$ level) is detected at the position
of IRS2. The flux densities of IRS1 and IRS2 are derived to be
140\,$\pm$\,28\,mJy and 2.8\,$\pm$\,2.1\,mJy, respectively. The
large-scale envelope detected in the 1.3\,mm single-dish map, with
a radius of $\sim$\,9000\,AU and a flux density of $\sim$\,3.7\,Jy
(B97), is also resolved out here. Positions and FHWM sizes of the
sources are listed in Table~2. The angular separation of
17$''$\,$\pm$\,1$''$ between IRS1 and IRS2 corresponds to a
projected linear separation of 3400\,$\pm$\,200\,AU at a distance
of 200\,pc. We also note that IRS1 is elongated
northwest-southeast and consists of two separate peaks in the
region enclosed by the 5\,$\sigma$ level (see Fig.\,1b). The main
peak is spatially coincident with the MIR source and the fainter
peak is located $\sim$\,2$''$ southeast of IRS1 (see below
$\S$\,3.3).

Assuming that the 3\,mm dust continuum emission is optically thin,
the hydrogen gas mass $M_{\rm H}$ = $M$(H) + 2\,$M$(H$_2$) in the
circumstellar envelope (excluding Helium) was calculated with the
same method as described in Launhardt \& Henning (1997). We adopt
an interstellar hydrogen-to-dust mass ratio of 110, and a dust
opacity $\kappa_{\rm 3mm} \approx 0.2$\,cm$^2$\,g$^{-1}$\ (using
$\kappa_{\rm 1.3mm} = 0.8$\,cm$^2$\,g$^{-1}$\ and
$\kappa~\propto~\nu^{1.8}$), a fairly typical value for dense
protostellar cores (Ossenkopf \& Henning 1994). Dust temperatures
are derived from SED fitting ($\S$\,4.1) and are listed in
Table~7. The derived hydrogen gas masses
(0.05\,$-$\,2.1\,$M_{\odot}$), together with mean volume densities
(0.4\,$-$\,2.6 $\times$ 10$^7$\,cm$^{-3}$) and column densities
(1.5\,$-$\,9.9 $\times$ 10$^{23}$ cm$^{-2}$), are listed in
Table~2. The resulting optical depths are $\tau_{\rm 3mm}$ $\sim$
0.4\,$-$\,3 $\times$ $10^{-3}$, thus justifying the optically thin
approximation.

\subsection{N$_2$H$^+$\,(1\,$-$\,0)}

N$_2$H$^+$ emission is detected from both CG\,30 and BHR\,71.
Figure~2a shows the velocity-integrated N$_2$H$^+$ intensity image
of CG\,30. Two cores, spatially associated with the 3\,mm dust
continuum sources, are seen. The northern core is elongated
east-west with a long ($\sim$\,20$''$) extension to the west,
along the direction of the protostellar jet HH\,120 (see
Fig.\,7c). The southern core is more compact and peaks at the
position of the dust continuum source. The mean radii of the
N$_{2}$H$^{+}$ cores (Table~4) were measured with the same method
described in Paper\,I. A larger-scale N$_2$H$^+$ cloud core,
detected in the Mopra single-dish map with a radius of
$\sim$\,8000\,AU (P.~Barnes et al. in prep.), peaks between the
two sub-cores (see Fig.\,2a) and is resolved out by the
interferometer (more than 90\% flux is missing).

Figure~2b shows the integrated N$_2$H$^+$ intensity image of
BHR\,71. Two cores are found to the east and west of IRS1 (see
below and discussion in $\S$\,4.4). We refer to these as BHR\,71E
and BHR\,71W, respectively. The two cores are elongated in the
north-south direction. Several smaller clumps are also seen north
and south of the two main cores, along both sides of the
large-scale CO outflow (see Fig.\,2b). For BHR\,71, the Mopra
N$_2$H$^+$ map (P.~Barnes et al. in prep.) again shows a
large-scale cloud core with one peak ($\sim$\,15$''$ offset IRS1),
and does not line up well with the two dust continuum sources (see
Fig.\,2b).

Figure~3 shows the N$_2$H$^{+}$ spectra at the peak positions of
CG\,30 and BHR\,71\footnote{During the observations towards
BHR\,71, the correlator was not well centered due to an
uncertainty in the Doppler correction calculation, resulting in
the N$_2$H$^{+}$ $JF_1F$\,=\,$101-012$ component not being
covered. However, the absence of this line component did not
affect our final results.}. The spectra were fitted using the
hyperfine program in CLASS. The fitting results, such as LSR
velocities ($V_{\rm LSR}$), intrinsic line width ($\triangle$$v$;
corrected for instrumental effects), total optical depths
($\tau_{\rm tot}$), and excitation temperatures ($T_{\rm ex}$),
are listed in Table~3.

Figure~4 shows the mean velocity fields of CG\,30 and BHR\,71,
derived from the N$_2$H$^{+}$ line maps with the fitting routine
described in Paper\,I. The jet/outflow information is also shown
in each map. In CG\,30, the southern core shows a well-ordered
velocity field, with gradient parallel to the outflow direction.
The northern core shows a more complicated velocity field, but the
gradient in the inner core is also parallel to the outflow
direction. In BHR\,71, there seems to be a general velocity
gradient across the two N$_2$H$^+$ cores, which is approximately
perpendicular to the axis of the large-scale CO outflow. This may
indicate that the two cores are actually part of one physical
structure associated with IRS1 (see discussion in $\S$\,4.4). A
least-squares fitting of the velocity gradients has been performed
using the routine described in Goodman et al. (1993). The results
are summarized in Table~5 and discussed in $\S$\,4.2.

Figure~5 shows the spatial distribution of N$_{2}$H$^{+}$ line
widths for both sources. The line widths are roughly constant
within the interiors of the cores, which is consistent with the
observational results in Paper\,I. The mean line widths were
derived through Gaussian fitting to the distribution of line
widths versus solid angle area in the maps (see Fig.\,6). We find
that the sub-cores in each object have roughly equal line width,
but the mean line width in CG\,30 ($\sim$\,0.5\,km\,s$^{-1}$) is
$\sim$\,1.7 times larger than that in BHR\,71
($\sim$\,0.3\,km\,s$^{-1}$).

Assuming that the observed N$_{2}$H$^{+}$ line widths are not
dominated by systematic gas motions, the virial mass of the cores
has been calculated as:
\begin{equation}
M_{\rm vir} = \frac{5}{8{\rm ln2}}\frac{R \triangle v_{\rm
ave}^2}{\alpha_{\rm vir} G},
\end{equation}
where $G$ is the gravitational constant, $R$ is the FWHM core
radius, and $\triangle v_{\rm ave}$ is the line width of the
emission from an ``average" particle with mass $m_{\rm ave}$ =
2.33 amu (assuming gas with 90\% H$_2$ and 10\% He). The
coefficient $\alpha_{\rm vir}$ = (1 $-$ $p$/3)/(1 $-$ 2$p$/5),
where $p$ is the power-law index of the density profile, is a
correction for deviations from constant density (see Williams et
al. 1994). In our calculations, we assume $p$ = 1.5 (see Andr\'{e}
et al. 2000) and use $\alpha_{\rm vir}$ = 1.25. $\triangle v_{\rm
ave}$ is derived from the observed spectra by
\begin{equation}
\triangle v_{\rm ave}^2 = \triangle v_{\rm obs}^2 + 8{\rm
ln2}\frac{kT_{\rm ex}}{m_{\rm H}}(\frac{1}{m_{\rm
ave}}-\frac{1}{m_{\rm obs}}),
\end{equation}
where $\triangle v_{\rm obs}$ is the observed mean line width of
N$_2$H$^+$ and $m_{\rm obs}$ is the mass of the emitting molecule
(here we use $m_{\rm N_2H^+}$ = 29 amu). We derive virial masses
between 0.1 and 0.6\,$M_\odot$. The results are listed in Table~4.

The N$_2$H$^+$ column density has been calculated independently
from the line intensity using the equation given by Benson et al.
(1998):
\begin{equation}
N({\rm N_{2}H^{+}}) = 3.3 \times 10^{11} \frac{\tau \triangle v
T_{\rm ex}}{1-e^{-4.47/T_{\rm ex}}}\,(\rm cm^{-2}),
\end{equation}
where $\tau$ is the total optical depth, $\triangle v$ is the
intrinsic line width in km\,s$^{-1}$, and $T_{\rm ex}$ is the
excitation temperature in K. The gas-phase N$_{2}$H$^{+}$ mass of
the core was then calculated from $M_{\rm
N_{2}H^{+}}$~$\approx$~$N(\rm N_2H^+)_{\rm
peak}$\,$\times$\,$m_{\rm
N_2H^+}$\,$\times$\,$d^{2}$\,$\times$\,$\Omega_{\rm FWHM}$, where
$d$ is the distance from the Sun and $\Omega_{\rm FWHM}$ is the
solid angle enclosed by the FWHM contours for each core.

Assuming that the gas mass and virial mass derived from the
N$_{2}$H$^{+}$ data are the same, we derived the average
fractional abundance of N$_{2}$H$^{+}$ in each core (see
Table\,4). The average value $\langle$$X(\rm N_{2}H^{+})$$\rangle$
$\sim$\,3.0\,$\times$ 10$^{-10}$ for CG\,30 and BHR\,71 is close
to the mean value found in Paper\,I ($\sim$ 3.3 $\times$
10$^{-10}$) for nine other protostellar cores.

\subsection{\emph{Spitzer} Images}

Figure~7 shows the $Spitzer$ images of CG\,30. The infrared
emission from CG\,30N and CG\,30S is detected at all IRAC bands
(3.6\,$\mu$m\,$-$\,8.0\,$\mu$m). Fig.\,7a shows a wide-field IRAC
band\,2 (4.5\,$\mu$m) image. Centered at CG\,30S is a highly
collimated bipolar jet, with P.A. $\sim$\,40$^\circ$. The knots in
the jet are labeled with the same numbers as in Hodapp \& Ladd
(1995). The most distant knot (No.\,8) is $\sim$ 90$''$ away from
CG\,30S. Assuming a typical jet speed of 100 km s$^{-1}$ (Reipurth
\& Bally 2001), an inclination angle of 90$^\circ$, and a distance
of 400\,pc, the dynamical age of the jet is estimated to be
$\sim$\,1700\,yr. CG\,30N appears to be the driving source of
HH\,120, which is $\sim$\,5$''$ in size and extends to the west.
Knot No.\,6, located to the east of CG\,30N, is probably ejected
by CG\,30N and part of the same outflow as HH\,120.

Figs.\,7b and 7c show enlarged views of the two sources, overlaid
with the contours from the ATCA 3\,mm dust continuum and
N$_2$H$^+$ images. The two infrared sources are spatially
coincident with the 3\,mm dust continuum and N$_2$H$^+$ sources.
However, when viewed in detail, CG\,30S is elongated at the
infrared bands and the continuum source is located at the apex of
the infrared emission, implying that the infrared emission at IRAC
bands from CG\,30S is due to scattered light in a cavity evacuated
by the jet/outflow. In contrast, CG\,30N shows a point-like
structure at all IRAC bands coincident with the circumstellar mm
dust emission peak, suggesting that the source is directly
detected at NIR wavelengths ($\lambda$ $<$ 5\,$\mu$m). The
N$_2$H$^+$ emission from CG\,30N spatially follows the direction
of the protostellar jet and the long extension to the west matches
exactly the HH\,120 flow (see Fig.\,7c), indicating that the jet
has a strong effect on the morphology of the N$_2$H$^+$ emission.

In the $Spitzer$ MIPS\,1 (24\,$\mu$m) image shown in Fig.\,7d,
CG\,30 is again resolved in two sources, but the emission is
dominated by CG\,30N and only weak emission is found at the
position of CG\,30S. In the MIPS\,2 (70\,$\mu$m) image (see
Fig.\,7e), the two sources are not fully resolved, but two peaks,
with flux ratio $\sim$ 2:1, can be clearly distinguished. Flux
densities of CG\,30N and CG\,30S in the IRAC and MIPS bands are
measured (see $\S$\,2.2) and listed in Table~6.

The $Spitzer$ images of BHR\,71 are shown in Fig.\,8, with the
same sequence as in Fig.\,7. The infrared emission from IRS1 and
IRS2 is detected at all IRAC bands. A large-scale ($\sim$ 160$''$
in length) bipolar jet, centered at IRS1 with a P.A. of
165$^\circ$, is seen in the IRAC images (Fig.\,8a). The northern
jet, spatially coincident with the red-shifted CO outflow, is
S-shaped, while the southern jet, containing the HH object HH\,321
(Corporon \& Reipurth 1997), shows a V-shaped structure at the
apex. This V-shaped structure may represent a conical cavity
evacuated by the successive bow-shocks traced by the infrared
emission (Fig.\,8a) and the blue-shifted CO outflow (see B97 and
Parise et al. 2006). Another bipolar jet, at P.A. $\sim$
30$^\circ$, is found with IRS2 being in the center. Its northwest
lobe, containing another HH object HH\,320 (Corporon \& Reipurth
1997), also shows a V-shaped structure at the apex and could be
explained in the same way.

IRS1 and IRS2 are spatially coincident with the dust continuum
sources detected with ATCA (see Fig.\,8b). We note that the IRS2
dust continuum source is located at the apex of the infrared
emission and could be explained in the same way as CG\,30S. The
elongated structure and secondary peak found in the ATCA dust
continuum image match the left wall of the outflow cavity,
suggesting they result from the jet/outflow action (for a similar
case, see Gueth et al. 2003). The N$_{2}$H$^{+}$ emission is
located on both sides of the large-scale CO outflow and basically
matches the wall of the cavity (see Fig.\,8c). At the MIPS\,1
band, BHR\,71 is barely resolved into two sources and the emission
is dominated by IRS1 (see Fig.\,8d). The MIPS\,2 image does not
resolve the two sources and the emission is peaked at the position
of IRS1\footnote{The offset ($\sim$\,3$''$) between the MIPS\,2
emission peak and the 3\,mm emission peak is much smaller than the
FWHM of MIPS\,2 PSF ($>$\,10$''$), and is not significant.} (see
Fig.\,8e). Flux densities of IRS1 and IRS2 are listed in Table~6.

\section{DISCUSSION}
\subsection{Spectral Energy Distributions and Evolutionary Stages}

Figure~9 shows the spectral energy distributions (SEDs) of CG\,30
N and S and BHR\,71 IRS1 and IRS2, based on the infrared (ISOCAM,
$Spitzer$, and IRAS), sub-mm (SCUBA, available only for CG\,30),
and mm (SEST and ATCA) observations. The NIR data of CG\,30N are
adopted from Persi et al. (1990). The SCUBA and SEST data for
CG\,30 are adopted from Henning et al. (2001) and Henning \&
Launhardt (1998), respectively. The ISOCAM and SEST data for
BHR\,71 are adopted from B01 and B97, respectively. Here we do not
explicitly list all flux values, but show graphically the SEDs.
Since IRAS observations could resolve neither CG\,30 nor BHR\,71,
flux ratios at the IRAS wavelengths of 10:1 (CG\,30N\,:\,CG\,30S)
and 20:1 (BHR\,71 IRS1\,:\,IRS2) were inferred from the $Spitzer$
and ATCA observations.

In order to derive luminosities and bolometric temperatures, we
first interpolated and then integrated the SEDs, always assuming
spherical symmetry. Interpolation between the flux densities was
done by a $\chi$$^2$ grey-body fit to all points at $\lambda$
$\geq$ 100\,$\mu$m\footnote{The 3\,mm points were ignored in the
fitting to CG\,30 and BHR\,71 IRS2 to give higher priority to the submm data,
resulting in much better fitting.}, using
\begin{equation}
S_{\nu} = B_{\nu}(T_{\rm d})(1 - e^{-\tau_{\nu}})\Omega,
\end{equation}
where $B_{\nu}$$(T_{\rm d})$ is the Planck function at frequency
$\nu$ and dust temperature $T_{\rm d}$, $\tau_{\nu}$ is the dust
optical depth as a function of frequency ${\tau} \propto
{\nu}^{1.8}$, and $\Omega$ is the solid angle of the source. A
simple logarithmic interpolation was used between all points at
$\lambda$ $\leq$ 100\,$\mu$m. The fitting results, such as dust
and bolometric temperatures, sub-mm ($\lambda$ $\geq$ 350\,$\mu$m)
and bolometric luminosities, are listed in Table~7.

Based on these results, we try to address the evolutionary stages
of CG\,30 and BHR\,71. A detailed definition and discussion for
early stellar evolutionary phases can be found in Andr\'{e} et al.
(2000) and Froebrich (2005). The $L_{\rm submm}$/$L_{\rm bol}$
ratios of all sources are $\gg$ 0.5\% (the standard boundary of
Class\,0 protostars, see Andr\'{e} et al. 2000)\footnote{Our
$L_{\rm submm}$/$L_{\rm bol}$ ratios are larger than those found
by Froebrich (2005). We attribute this to the fact that Frobrich
(2005) assumed the two objects were single cores, but we resolved
them as binaries. Furthermore, we have more data points at submm
wavelengths (for CG\,30) and high-resolution interferometric data
points at 3\,mm, which were all not available to Froebrich
(2005).} and the four sources each drive a bipolar jet (see
$\S$\,3.3). However, the bolometric temperature of CG\,30N is
$\sim$\,100\,K and the object is also directly detected at NIR
wavelengths, suggesting CG\,30N is a Class\,I young stellar
object. In contrast, the low bolometric temperature (37\,K) of
CG\,30S suggests it is a Class\,0 protostar. In BHR\,71, both IRS1
and IRS2 have bolometric temperatures less than 70\,K (see
Table~7). Nevertheless, IRS1 might be directly detected at NIR
wavelengths (see Fig.\,8a), suggesting that it is a transition
object between Class\,0 and I, while IRS2 could be a Class\,0
protostar.

It must be noted that the analysis above does not take into account inclination effects: considering a protostar embedded in a circumstellar disk/envelope, its infrared emission could be detected through the outflow cavity when the system is face-on, but is not seen when it is edge-on. In BHR\,71, the bipolar CO outflow powered by IRS1 is lying roughly in the plane of sky, implying the latter case; the bipolar outflow driven by IRS2 appears to favor the same situation (see Parise et al. 2006). In CG\,30, however, the relative inclinations could not be easily distinguished because the information about molecular outflows is still missing. The SED-based classification discussed above thus might be a result of both evolutionary stage and inclination. In particular in CG\,30 we cannot disentangle the two effects. It is well possible that the different SEDs (and bolometric temperatures) reflect actually inclination effects rather than a difference in evolutionary stage.

\subsection{Gas Kinematics}


The thermal contribution to the N$_{2}$H$^{+}$ line width is
calculated by $\triangle v_{\rm th}^2 = {\rm 8ln2}\frac{kT_{\rm
K}}{m_{\rm obs}}$, where $k$ is the Boltzmann constant, $T_{\rm
K}$ is the kinetic gas temperature, and $m_{\rm obs}$ is the mass
of the observed molecule. Assuming that at the high densities of
$>$\,10$^{6}$\,cm$^{-3}$ (see Table~2) the kinetic gas temperature
is equal to the dust temperature derived in $\S$\,4.1
($\sim$\,20\,K), the non-thermal contributions to the line widths
($\triangle v_{\rm NT} = \sqrt{\triangle v_{\rm mean}^2 -
\triangle v_{\rm th}^2}$) were then calculated to be
$\sim$\,0.5\,km\,s$^{-1}$ in CG\,30 and $\sim$\,0.2\,km\,s$^{-1}$
in BHR\,71 (see Table~3). These non-thermal line widths suggest
that turbulence, the main contribution to the non-thermal line
width (Goodman et al. 1998), cannot be ignored in the protostellar
cores. On the other hand, the thermal line width of an ``average"
particle of mass 2.33\,$m_{\rm H}$ (assuming gas with 90\% H$_2$
and 10\% He), which represents the local sound speed, is
$\sim$\,0.62\,km\,s$^{-1}$ at 20\,K. The derived non-thermal
contributions to the N$_{2}$H$^{+}$ line width in both CG\,30 and
BHR\,71 are smaller than this local sound speed (i.e., the
turbulent motion is subsonic). We also note that the mean line
widths derived for BHR\,71 ($\sim$\,0.3\,km\,s$^{-1}$) are three
times smaller than measured by single-dish observations in
Mardones et al. (1997; $\sim$\,0.9\,km\,s$^{-1}$). Taking into
account the systematic velocity variation across the core
($\sim$\,0.3\,km\,s$^{-1}$; see Table~5), the combined line width
in our maps is still smaller than the result from single-dish
observations. It means that high-level (supersonic) turbulence
occurs mainly in the extended envelope which is resolved out by
the interferometer, but the inner core is much more ``quiescent".
This is consistent with what we found in Paper\,I, namely that
non-thermal motions are quickly damped from large-scale to smaller
inner cores (see e.g., Fuller \& Myers 1992).


The velocity fields of CG\,30N, CG\,30S, and BHR\,71 show
systematic velocity gradients (see Fig.\,4). As discussed in
Paper\,I, systematic velocity gradients are usually dominated by
either rotation or outflow. In CG\,30, the gradients in both cores
are parallel to the jets. Although there is no molecular outflow
information available yet for CG\,30, these gradients are likely
the results of outflows and we treat them as upper limits of
underlying rotation velocity gradients. In BHR\,71, the velocity
gradient across the two N$_{2}$H$^{+}$ cores is roughly
perpendicular to the axis of the large-scale CO outflow and could
be explained by rotation. [Here we assume that the two cores are
associated with IRS1 (see $\S$\,4.4).] The velocity gradients
measured in CG\,30N, CG\,30S, and BHR\,71 are $<$\,24.4,
$<$\,17.5, and 7.8\,$\pm$\,0.5\,km\,s$^{-1}$\,pc$^{-1}$,
respectively (see Table~5). Both the velocity gradient in BHR\,71
as well as the upper limits for CG\,30N and CG\,30S are consistent
with those found in Paper\,I.

Assuming that the velocity gradients summarized in Table~5 are due
to core rotation, the specific angular momentum $J/M$ of the
objects was calculated with the following equation:
\begin{equation}
J/M = \alpha_{\rm rot}\omega R^2 = \frac{2}{3} \frac{3 - p}{5 - p}
\frac{g}{sini}R^2 \approx \frac{2}{7}\,gR^2,
\end{equation}
where the coefficient $\alpha_{\rm rot}$ = $\frac{2}{3}$ $\frac{3
- p}{5 - p}$, $p$ is the power-law index of the radial density
profile ($p$ = 1.5; see $\S$\,3.2), $g$ is the velocity gradient,
and $i$ is the inclination angle to the line of sight direction
(here we assume $sin\,i$ = 1). The derived $J/M$ for CG\,30N,
CG\,30S, and BHR\,71 are listed in Table~5. It should be noted
that for CG\,30 we derive only upper limits. The ratio of
rotational energy to the gravitational potential energy was
calculated by $\beta_{\rm rot}$ = $\frac{E_{\rm rot}}{E_{\rm
grav}}$ $\approx$ 0.19\,$\frac{g^2R^3}{GM}$, where $E_{\rm rot}$ =
$\frac{1}{2}$ $I$ $\omega^2$ = $\frac{1}{2}$ $\alpha_{\rm rot}$
$MR^2$ $\omega^2$ and $E_{\rm grav}$ = $\frac{3}{5}$ $\alpha_{\rm
vir}GM^2/R$ (the masses and radii used in the equations are virial
masses and radii listed in Table~4). The estimated $\beta_{\rm
rot}$ values for CG\,30N, CG\,30S, and BHR\,71 are $<$\,0.019,
$<$\,0.014, and $\sim$\,0.020, respectively.

\subsection{How did the cores fragment?}

Recent numerical simulations and observations support the
hypothesis that the fragmentation of molecular cloud cores is the
main mechanism for the formation of binary/multiple stellar
systems, although the exact {\it when}, {\it where}, {\it why},
and {\it how} are still under debate (see reviews by Bodenheimer
et al. 2000, Tohline 2002, and Goodwin et al. 2007). In this
section, we try to examine the origin of the sub-cores in both
CG\,30 and BHR\,71, i.e., whether they formed by initial cloud
fragmentation prior to protostellar collapse or by prompt
rotational fragmentation of a single core after the initial
collapse.

In CG\,30, our previous single-dish submm maps have shown a
large-scale hourglass-shaped common envelope around the two
sub-cores (Henning et al. 2001; see Fig.\,1a). The separation
between the sub-cores is $\sim$\,8700\,AU, which is roughly two
times the typical Jeans length [$R_{\rm Jeans}$ =
0.19\,pc\,($\frac{T}{10K}$)$^{\frac{1}{2}}$\,($\frac{n_{\rm
H_2}}{10^4cm^{-3}}$)$^{-\frac{1}{2}}$; see Stahler \& Palla 2004]
in prestellar cores ($\sim$\,4000\,AU at $T$ = 10\,K and $n_{\rm
H_2}$ = 10$^6$\,cm$^{-3}$). The radial velocity difference between
the two sub-cores is $\sim$\,0.16\,km\,s$^{-1}$ (see
Table~3). If we assume that the total binary mass is
1.4\,$\times$\,1.36\,$M_\odot$ (see Table~2; the factor 1.36 accounting for He and heavier elements) and the orbit is perpendicular to the
plane of sky, the orbit velocity difference in a bound system with
the separation of 8700\,AU should be $\sim$\,0.44\,km\,s$^{-1}$,
about three times larger than the observed value. Furthermore, from
this observed velocity difference, we estimate the $\beta_{\rm
rot}$ of $\sim$ 0.008 for the large-scale cloud core which
contains the two sub-cores (radius $\sim$\,8000\,AU; see
Fig.\,2a). This $\beta_{\rm rot}$ is less than the typical
boundary suggested by a series of numerical simulations (see e.g.,
Boss 1999 and Machida et al. 2005) for rotational fragmentation.
Based on the morphology and velocity structure, we suggest that
the two sub-cores in CG\,30 were formed by initial
fragmentation\footnote{The basic idea of this initial
fragmentation is that the collapse is initiated in a large-scale
molecular cloud core which contains multiple Jeans masses in a
weakly condensed configuration, e.g., a prolate or filamentary
Gaussian distribution with several Jeans masses along the long
axis and one Jeans mass across the short axis; with some initial
angular momentum, provided by either slow rotation (Bonnell et al.
1991) or turbulence (Goodwin et al. 2007), the large cloud core
fragments at Jeans scale into several dense cores, in which the
separate protostellar collapse then starts and proceeds more
quickly than across the whole structure (see e.g., Mundy et al.
2001 and reference therein).} of a large-scale filamentary
prestellar core.

In BHR\,71, the two sub-cores have a separation of
$\sim$\,3400\,AU (less than the typical Jeans length) and are also
surrounded by a large common envelope (Fig.\,1b). Unfortunately,
the observed velocity structure is mainly associated with IRS1 and
kinematic information of IRS2 is missing. Here we can only
speculate on the basis of separation that the two sub-cores could
be formed by prompt rotational fragmentation of a collapsing
protostellar core.

Numerical simulations also predict that the material collapses
along the magnetic field lines while the fragmentation occurs in a
plane perpendicular to the magnetic field. This is supported by
our previous submm polarimetric observations towards CG\,30
(Henning et al. 2001). In contrast to a simple assumption that the
angular momenta of two components will be parallel in the
fragmentation, we find that the outflows, and hence the angular
momenta (assumed to be parallel to the outflows), of the sub-cores
are not aligned, neither in CG\,30, nor in BHR\,71. This
phenomenon is also found in other binary protostars studied
recently, like e.g., CB\,230 and L\,723 (Launhardt 2004; Launhardt
et al., in prep.). This could mean that during core fragmentation
the initial angular momentum is not evenly (in value and
direction) divided between the sub-cores, although the mean
direction is preserved all the time.

\subsection{N$_{2}$H$^{+}$ vs. Dust vs. CO}

From our observations towards Class\,0 protostars conducted at
OVRO (Paper\,I), ATCA (this work), and IRAM-PdBI (Chen et al., in
prep.), we find that in most objects the mm continuum source lies
within the half maximum level of the N$_2$H$^{+}$ emission. This
good general agreement indicates that N$_2$H$^{+}$ is spatially
associated with thermal dust in dense protostellar cores and
cannot be significantly depleted like, e.g., CO and CS (see Bergin
et al. 2001 and Caselli et al. 1999).

Fig.\,10 shows that the dust mass (converted into hydrogen gas
mass) is in general correlated with both the N$_2$H$^{+}$ gas mass
and the virial mass (both derived from the N$_2$H$^{+}$ emission).
However, there is a significant scatter in both correlations,
indicating that the agreement holds only within a factor of 2 to
2.5 (1\,$\sigma$ scatter). This could be due to the fact that the
mm dust continuum emission traces mainly the dense structures
(e.g., inner envelope or disk), while N$_2$H$^{+}$ emission traces
the larger-scale envelope (see e.g., Figs.\,1a\,\&\,2a), and hence
reflects different masses on different scales. We also note that
the N$_2$H$^{+}$ gas mass depends on the specific source
morphology and chemistry since it is quickly destroyed where CO is
released from dust grain into the gas phase (see below). The
estimated virial mass has also significant uncertainties because
several sources are driving bipolar outflows and are probably no
longer in virial equilibrium.

On the other hand, we also find that in most objects the
morphology of the N$_{2}$H$^{+}$ emission is directly related to
the jet/outflow actions. For example, in BHR\,71, two
N$_{2}$H$^{+}$ cores, located to the east and west of the
outflow-driving source IRS1, are rotating perpendicular to the
outflow axis, and there is no N$_{2}$H$^{+}$ emission detected at
the origin and along the large-scale CO outflow. These features
suggest a large N$_{2}$H$^{+}$ hole has been formed and the two
cores may be the remnant of a N$_{2}$H$^{+}$ envelope\footnote{For
similar cases see low-mass protostars L\,483 (J$\o$rgensen 2004)
and IRAM\,04191 (Belloche \& Andr\'{e} 2004).}. It is likely that
a large amount of N$_{2}$H$^{+}$ in the way of the outflow has
been depleted by CO molecules, which is one of the main destroyers
of N$_{2}$H$^{+}$ in the gas phase (Aikawa et al. 2001). For this
reason, we think that the emission at the position of IRS2 (see
Fig.\,2b) is part of the structure around IRS1 and does not
originate from IRS2.

Based on the observational results, we speculate that there are
three stages of the interaction between N$_{2}$H$^{+}$ and
jets/outflows. (1) When jets are ejected from a protostar,
N$_{2}$H$^{+}$ molecules in the envelope are entrained and show a
jet-like morphology in the images, like, e.g., L723 VLA2
(Paper\,I) and CG\,30N (this work). (2) Molecular outflows,
following the jets, release CO from grain surfaces back into the
gas phase and start destroying the N$_{2}$H$^{+}$ molecules on the
way, leading to the observed hourglass structure perpendicular to
the CO outflow axis, like e.g., IRAS\,03282+3035,
IRAS\,04166+2706, and CB\,224 (see Paper\,I). (3) Large
N$_{2}$H$^{+}$ holes form in the envelopes, like e.g., in BHR\,71
IRS1 (this work), L\,483 (J$\o$rgensen 2004), and IRAM\,04191
(Belloche \& Andr\'{e} 2004). However, there seems to be no clear
correlation between this N$_{2}$H$^{+}$/jet scenario and standard
evolutionary scenario from Class\,0 to Class\,I. For example,
IRAM\,04191 is a young Class\,0 protostar but appears in the last
stage, while CG\,30N is a Class\,I object but appears in the first
stage. We speculate that the appearance of the N$_{2}$H$^{+}$
emission is strongly affected by outflow-envelope interaction,
which depends on the specific envelope morphology and source
multiplicity properties.

\section{SUMMARY}

We have presented ATCA and $Spitzer$ observations of the two
isolated protostellar double cores CG\,30 and BHR\,71 in the
southern sky. The main results of this work are summarized as
follows:

(1) The 3\,mm dust continuum image of CG\,30 resolves two compact
sources with a separation of $\sim$ 21$''$ (8400\,AU). In BHR\,71,
one strong dust continuum source is detected at the position of
mid-infrared source IRS1, while only weak emission is detected
from the secondary mid-infrared source IRS2. The separation
between IRS1 and IRS2 is $\sim$ 17$''$ (3400\,AU). Assuming
optically thin dust emission, we derive hydrogen gas masses of
1.1\,$M_\odot$ and 0.33\,$M_\odot$ for northern and southern
sources in CG\,30, and 2.1\,$M_\odot$ and 0.05\,$M_\odot$ for IRS1
and IRS2 sources in BHR\,71.

(2) N$_{2}$H$^{+}$\,(1\,$-$\,0) emission is detected in both
CG\,30 and BHR\,71. In CG\,30, the two dust continuum sources are
directly associated with N$_{2}$H$^{+}$ cores. In BHR\,71, two
N$_{2}$H$^{+}$ cores are around the primary dust continuum source,
probably part of one large envelope, but no N$_{2}$H$^{+}$ is
detected at the position of the dust source. The secondary IR
source is not detected in N$_{2}$H$^{+}$.

(3) The excitation temperatures of the N$_{2}$H$^{+}$ line are
4.7\,$-$\,6.8\,K for CG\,30 and 3.9\,$-$\,4.4\,K for BHR\,71. The
FWHM radii of N$_{2}$H$^{+}$ cores range from 730 to 1700\,AU. The
average fractional abundances of N$_{2}$H$^{+}$, derived from the
ratio of N$_{2}$H$^{+}$ gas mass to virial mass, is
$\sim$\,3.0\,$\times$\,10$^{-10}$, which is consistent with the
results obtained in our previous study of the cores in northern
sky. The observed mean N$_{2}$H$^{+}$ line widths are
$\sim$\,0.5\,km\,s$^{-1}$ for CG\,30 and $\sim$\,0.3\,km\,s$^{-1}$
for BHR\,71. The line widths are roughly constant within the
interiors of the cores and large line widths only occur at the
edges of the cores. The derived virial masses of the
N$_{2}$H$^{+}$ cores range from 0.1 to 0.6\,$M_\odot$.

(4) We derive the N$_{2}$H$^{+}$ radial velocity fields for CG\,30
and BHR\,71. The two N$_{2}$H$^{+}$ cores in CG\,30 show
systematic velocity gradients of
$\sim$\,24.4\,km\,s$^{-1}$\,pc$^{-1}$ and
$\sim$\,17.8\,km\,s$^{-1}$\,pc$^{-1}$ that are parallel to the
outflow directions and could be affected by the outflows. In
BHR\,71, a systematic velocity gradient of
$\sim$\,7.8\,km\,s$^{-1}$\,pc$^{-1}$ across the two cores is
perpendicular to the large-scale outflow and could be explained by
rotation.

(5) Assuming that the observed velocity gradients are due to core
rotation (if perpendicular to outflow) or place an upper limit to
rotation (if parallel to outflow), we estimate specific angular
momenta of $<$\,0.30, $<$\,0.35, and $\sim$\,0.51 $\times$
10$^{-3}$\,km\,s$^{-1}$\,pc for CG\,30N, CG\,30S, and BHR\,71,
respectively. The ratios for the rotational energy to the
gravitational potential energy for CG\,30N, CG\,30S, and BHR\,71
are estimated to be $<$\,0.019, $<$\,0.014, and $\sim$\,0.020,
respectively.

(6) Infrared emission from both sub-cores in both CG\,30 and
BHR\,71 is detected at $Spitzer$ IRAC bands and MIPS bands. Each
source is driving its own outflow, as seen in the shock-excited
4.5\,$\mu$m infrared images. CG\,30N is associated with a
Herbig-Haro flow, while the southern source is driving a large
bipolar jet. In BHR\,71, both IRS1 and IRS2 are associated with
Herbig-Haro objects and driving bipolar jets which coincide
spatially with the CO outflows.

(7) By fitting the spectral energy distributions, we derive the
dust temperature, bolometric temperature, and bolometric
luminosity of the sources. We find that CG\,30N is a Class\,I
object while the southern source is a Class\,0 protostar. In
BHR\,71, the properties of IRS1 resemble a Class\,0/I transition
object, while IRS2 is a Class\,0 protostar. We speculate that the
sources may nevertheless be coeval but that this evolutionary
discrepancy is due to the effects of ralative inclinations.

(8) Based on the morphologies and velocity structures, we suggest
that the double cores in CG\,30 were formed by initial
fragmentation of a filamentary prestellar core, while BHR\,71 may
originate from rotational fragmentation of a single collapsing
protostellar core. We also find that the angular momenta of the
sub-cores are not aligned in either pair of sources.

(9) Our observations conducted at OVRO and ATCA show a close
correlation between thermal dust emission and N$_{2}$H$^{+}$. The
N$_{2}$H$^{+}$ emission in most sources is spatially associated
and quantitatively correlated with the dust continuum emission.
However, we also find a strong relationship between the morphology
of the N$_{2}$H$^{+}$ emission and the jet/outflow actions.
Outflows first seem to entrain N$_{2}$H$^{+}$ and then gradually
destroy it, which leads to the observed jet-like, hourglass-shaped
intensity maps and N$_{2}$H$^{+}$ hole.

\acknowledgments

We thank the anonymous referee for many helpful comments and
suggestions. The Australia Telescope Compact Array is part of the
Australia Telescope, which is funded by the Commonwealth of
Australia for operation as a national facility managed by CSIRO.
We thank the ATCA staff for technical support during the
observations. We also thank A.~Goodman for fruitful discussions
and providing the VFIT routine.

\clearpage
\begin{deluxetable}{lcccccc}
\tabletypesize{\scriptsize} \tablecaption{\footnotesize Target
list and summary of observations\label{tbl-1}} \tablewidth{0pt}
\tablehead{\colhead{Object} &\colhead{IRAS}&\colhead{R.A. \& Dec.
(J2000)$^{a}$} &\colhead{Distance}&\colhead{Array}
&\colhead{HPBW$^{b}$} &\colhead{rms$^c$}\\
\colhead{Name}&\colhead{Source}&\colhead{[h\,:\,m\,:\,s,
$^{\circ}:\,':\,''$]}&\colhead{[pc]}&\colhead{configuration}
&\colhead{[arcsec]} &\colhead{[mJy/beam]}}\startdata

CG\,30   & 08076$-$3556 & 08:09:33.0, $-$36:05:01.00  &400& H75+H168   & 4.6$\times$3.3/4.6$\times$3.2 & 65/0.5\\
BHR\,71  & 11590$-$6452 & 12:01:36.5, $-$65:08:49.49  &200& H75+H168   & 3.6$\times$2.9/3.9$\times$3.1 & 20/2.0\\

\enddata
\tablenotetext{a}{Reference position for figures and tables in the
paper (except $Spitzer$ images).} \tablenotetext{b}{Synthesized
FWHM beam sizes at N$_2$H$^+$(1\,$-$\,0) line\,/\,3\,mm dust
continuum with robust weighting 1.}\tablenotetext{c}{1\,$\sigma$
noises at N$_2$H$^+$(1\,$-$\,0) line\,/\,3\,mm dust continuum.}
\end{deluxetable}

\begin{deluxetable}{lllcccccc}
\tabletypesize{\scriptsize} \tablecaption{3\,mm dust continuum
results for CG\,30 and BHR\,71\label{tbl-2}} \tablewidth{0pt}
\tablehead{\colhead{Source}&\colhead{R.A.$^{a}$}&\colhead{Dec.$^{a}$}&
\colhead{$S_{\nu}$} & \multicolumn{2}{c}{FWHM sizes$^{a}$} &
\colhead{$M_{\rm H}$} &
\colhead{$\langle n_{\rm H} \rangle$$^b$}&\colhead{$N_{\rm H}$$^c$}\\
\cline{5-6}\colhead{}& \colhead{(J2000)}& \colhead{(J2000)} &
\colhead{[mJy]} & \colhead{maj.$\times$min.}& \colhead{P.A.} &
\colhead{[$M_{\odot}$]} & \colhead{[$\rm \times 10^7 cm^{-3}$]} &
\colhead{[$\rm \times 10^{23} cm^{-2}$]}}

\startdata
CG\,30N           & 08:09:33.12 & $-$36:04:58.12 & 15.8$\pm$3.2  & 5\farcs1$\times$3\farcs1 &89$\pm$7\degr     & 1.10$\pm$0.26  & 1.11 & 4.51\\
CG\,30S           & 08:09:32.67 & $-$36:05:19.09 &  6.0$\pm$1.3  & 4\farcs8$\times$3\farcs1 &74$\pm$15\degr    & 0.33$\pm$0.10  & 0.37 & 1.45\\
BHR\,71 IRS1      & 12:01:36.81 & $-$65:08:49.22 &  140$\pm$28   & 7\farcs8$\times$7\farcs1 &$-$73$\pm$20\degr & 2.12$\pm$0.41  & 2.64 & 9.94\\
BHR\,71 IRS2      & 12:01:34.09 & $-$65:08:47.36 &  2.8$\pm$2.1  & 2\farcs6$\times$2\farcs1 &76$\pm$40\degr    & 0.05$\pm$0.02  & 2.18 & 2.58\\
\enddata

\tablenotetext{a}{Center position and FWHM sizes of the continuum
sources derived from Gaussian $uv$ plane fitting.}
\tablenotetext{b}{Assuming a spherical morphology for the objects,
the mean volume density of hydrogen atoms $n_{\rm H}$ = $n$(H) +
2$n$(H$_2$) was calculated by $n_{\rm H}$ = $M_{\rm H}$/$m_{\rm
H}$$V$, with $V$ $\sim$ $\pi$/6($\theta_{\rm S}D$)$^3$ being
volume.}\tablenotetext{c}{The hydrogen column density $N_{\rm H}$
= $N$(H) + 2\,$N$(H$_2$) was derived from the flux densities by
$N_{\rm H}$ = $\frac{S_{\nu}}{\kappa_{\rm d}(\nu )\,\Omega_{\rm
S}\,B_{\nu}(\nu,T_{\rm d})}$\,$\frac{1}{m_{\rm H}}$
$\left(\frac{M_{\rm H}}{M_{\rm d}}\right)$, where $\Omega_{\rm S}$
is the solid angle of the objects and $m_{\rm H}$ is the proton
mass.}
\end{deluxetable}


\begin{deluxetable}{lrccccc}
\tabletypesize{\scriptsize}\tablecaption{\footnotesize Observing
parameters from N$_2$H$^+$\,(1\,$-$\,0) spectra
fitting\label{tbl-3}} \tablewidth{0pt} \tablehead{\colhead{}&
\colhead{$V_{\rm LSR}$$^{a}$} & \colhead{$\triangle$$v$$^{a}$} &
\colhead{$\tau_{\rm tot}$$^{a}$} & \colhead{$T_{\rm ex}$$^{a}$} &
\colhead{$\triangle$$v_{\rm mean}$$^b$}&
\colhead{$\triangle$$v_{\rm NT}$$^c$}\\
\colhead{Source} & \colhead{[km s$^{-1}$]} & \colhead{[km
s$^{-1}$]} & \colhead{} & \colhead{[K]} & \colhead{[km s$^{-1}$]}&
\colhead{[km s$^{-1}$]}} \startdata

CG\,30\,N       & 6.64$\pm$0.02      & 0.53$\pm$0.03   &  1.0$\pm$0.1   & 4.66$\pm$0.09 & 0.51$\pm$0.01 &0.47\\
CG\,30\,S       & 6.48$\pm$0.01      & 0.52$\pm$0.02   &  1.4$\pm$0.1   & 6.81$\pm$0.05 & 0.52$\pm$0.01 &0.48\\
BHR\,71\,E      & $-$4.35$\pm$0.02   & 0.44$\pm$0.10   &  1.9$\pm$0.4   & 4.44$\pm$0.13 & 0.28$\pm$0.01 &0.20\\
BHR\,71\,W      & $-$4.42$\pm$0.02   & 0.38$\pm$0.06   &  2.5$\pm$0.2   & 3.91$\pm$0.05 & 0.33$\pm$0.01 &0.27\\

\enddata
\tablenotetext{a}{Value at the intensity peak. The error
represents 1\,$\sigma$ error in the hyperfine
fitting.}\tablenotetext{b}{Mean line width obtained through
Gaussian fitting to the distribution of line widths versus solid
angle areas.}\tablenotetext{c}{Non-thermal line width at the given
dust temperature (see Table~7).}
\end{deluxetable}

\begin{deluxetable}{lcccccc}
\tabletypesize{\scriptsize} \tablecaption{\footnotesize Volume
size, density, and mass of N$_2$H$^+$ cores\label{tbl-4}}
\tablewidth{0pt} \tablehead{\colhead{Source} &\colhead{$R$}
&\colhead{$M_{\rm vir}$}
&\colhead{$n_{\rm vir}$} &\colhead{$N(\rm N_{2}H^{+})$} &\colhead{$M_{\rm N_2H^+}$}&\colhead{$X(\rm N_2H^+)$}\\
\colhead{}&\colhead{[AU]}&\colhead{[$M_{\odot}$]}&\colhead{[$\times$10$^6$
cm$^{-3}$]}&\colhead{[$\times$10$^{12}$
cm$^{-2}$]}&\colhead{[$\times$10$^{-10}$
$M_{\odot}$]}&\colhead{[$\times$10$^{-10}$]}}\startdata
CG\,30\,N           & 1300 & 0.38 &  6.2 & 1.29  & 0.51 & 1.83  \\
CG\,30\,S           & 1650 & 0.55 &  4.5 & 3.42  & 1.94 & 4.85  \\
BHR\,71\,E          &  960 & 0.13 &  5.3 & 1.95  & 0.34 & 3.63  \\
BHR\,71\,W          &  730 & 0.11 & 10.3 & 1.87  & 0.21 & 2.52  \\
BHR\,71 IRS1$^a$    & 3000 & 0.49 &  0.7 & 1.90  & 3.07 & 8.55  \\
\enddata
\tablenotetext{a}{Assuming that the two N$_2$H$^+$ cores found in
BHR\,71 are part of a physical structure around IRS1 with a radius
of $\sim$\,3000\,AU (see Fig.\,2b) and a mean line width of
0.33\,km\,s$^{-1}$ (see Table~3).}
\end{deluxetable}


\begin{deluxetable}{lcccccc}
\tabletypesize{\scriptsize} \tablecaption{\footnotesize Velocity
gradients and specific angular momentum\label{tbl-5}}
\tablewidth{0pt} \tablehead{\colhead{}&\colhead{mean
velocity}&\colhead{$g$}&\colhead{$\Theta_{g}^a$}&\colhead{$g_{r}$}&\colhead{$J/M$}&\colhead{$\beta_{\rm rot}$}\\
\colhead{Source}&\colhead{[km s$^{-1}$]}&\colhead{[km s$^{-1}$
pc$^{-1}$]}&\colhead{[degree]}&\colhead{[km
s$^{-1}$]}&\colhead{[$\times$10$^{-3}$ km s$^{-1}$
pc]}&\colhead{}}\startdata
CG\,30\,N          & 6.62    & $<$24.4$\pm$0.2     & $-$79.7$\pm$0.4    & 0.32  & $<$ 0.30    & $<$0.019  \\
CG\,30\,S          & 6.45    & $<$17.8$\pm$0.2     & 33.9$\pm$0.3       & 0.29  & $<$ 0.35    & $<$0.014  \\
BHR\,71 IRS1$^b$   & $-$4.39 & $\sim$7.8$\pm$0.5   & $-$104$\pm$2.0     & 0.23  & $\sim$ 0.51 & $\sim$0.020  \\
\enddata
\tablenotetext{a}{East of north in the direction of increasing
velocity} \tablenotetext{b}{The same assumption as in Table~4.}
\end{deluxetable}

\begin{deluxetable}{lllcccccc}
\tabletypesize{\scriptsize} \tablecaption{\footnotesize $Spitzer$
flux densities of CG\,30 and BHR\,71$^a$ \label{tbl-6}}
\tablewidth{0pt} \tablehead{\colhead{} &\colhead{R.A.$^b$}
&\colhead{Dec.$^b$} &\colhead{$S(3.6\,\mu m)$}
&\colhead{$S(4.5\,\mu m)$} &\colhead{$S(5.8\,\mu m)$}
&\colhead{$S(8.0\,\mu m)$} &\colhead{$S(24\,\mu m)$}&\colhead{$S(70\,\mu m)$}\\
\colhead{Source} &\colhead{(J2000)} &\colhead{(J2000)}
&\colhead{[mJy]} &\colhead{[mJy]} &\colhead{[mJy]}
&\colhead{[mJy]} &\colhead{[mJy]}&\colhead{[mJy]}}\startdata

CG\,30N           & 08:09:33.20 & $-$36:04:58.17 & 55.7$\pm$1.4 & 123.4$\pm$2.1 & 256.1$\pm$3.0 & 395.8$\pm$3.8 & 3400$\pm$100 & 8700$\pm$430\\
CG\,30S           & 08:09:32.68 & $-$36:05:20.38 & 6.7$\pm$0.5  & 16.8$\pm$0.8  & 19.6$\pm$0.8  & 9.6$\pm$0.6   & 50$\pm$5     & 4200$\pm$340\\
BHR\,71 IRS1      & 12:01:36.57 & $-$65:08:49.52 & 32.4$\pm$1.1 & 82.4$\pm$1.7  & 123.3$\pm$2.1 & 210.2$\pm$2.8 & 5000$\pm$300 & 84000$\pm$800\\
BHR\,71 IRS2      & 12:01:34.05 & $-$65:08:47.03 & 4.5$\pm$0.4  & 12.4$\pm$0.7  & 15.4$\pm$0.7  & 9.3$\pm$0.6   & 90$\pm$30    & $--$\\

\enddata
\tablenotetext{a}{Flux densities in the IRAC and MIPS bands were
measured using IRAF APPHOT and GILDAS, respectively (see
$\S$\,2.2).} \tablenotetext{b}{Peak position of infrared sources
measured at the IRAC band 3 (5.8\,$\mu$m).}
\end{deluxetable}


\begin{deluxetable}{lcccccc}
\tabletypesize{\scriptsize} \tablecaption{\footnotesize Fitting
results of the spectral energy distribution} \label{table:7}
\tablewidth{0pt}
\tablehead{\colhead{Source}&\colhead{$T_{\rm dust}$}&\colhead{$T_{\rm bol}$}&\colhead{$L_{\rm bol}$}
&\colhead{$L_{\rm submm}$}&\colhead{$L_{\rm submm}$/$L_{\rm bol}$}&\colhead{Classification}\\
\colhead{} &\colhead{[K]} &\colhead{[K]}
&\colhead{[$L_\odot$]}&\colhead{[$L_\odot$]}&\colhead{[\%]}&\colhead{}}\startdata
CG\,30N         & 22 & 102    & 13.6$\pm$0.8 & 0.49$\pm$0.10    &3.6 & Class\,I\\
CG\,30S         & 27 & 37     &  4.3$\pm$0.5 & 0.32$\pm$0.05    &7.4 & Class\,0\\
BHR\,71 IRS1    & 25 & 44     & 13.5$\pm$1.0 & 0.49$\pm$0.05    &3.6 & Class\,0/I\\
BHR\,71 IRS2    & 26 & 58     &  0.5$\pm$0.1 & 0.02$\pm$0.01    &3.4 & Class\,0\\
\enddata
\end{deluxetable}


\clearpage

\begin{figure*}
\begin{center}
\includegraphics[width=18cm,angle=0]{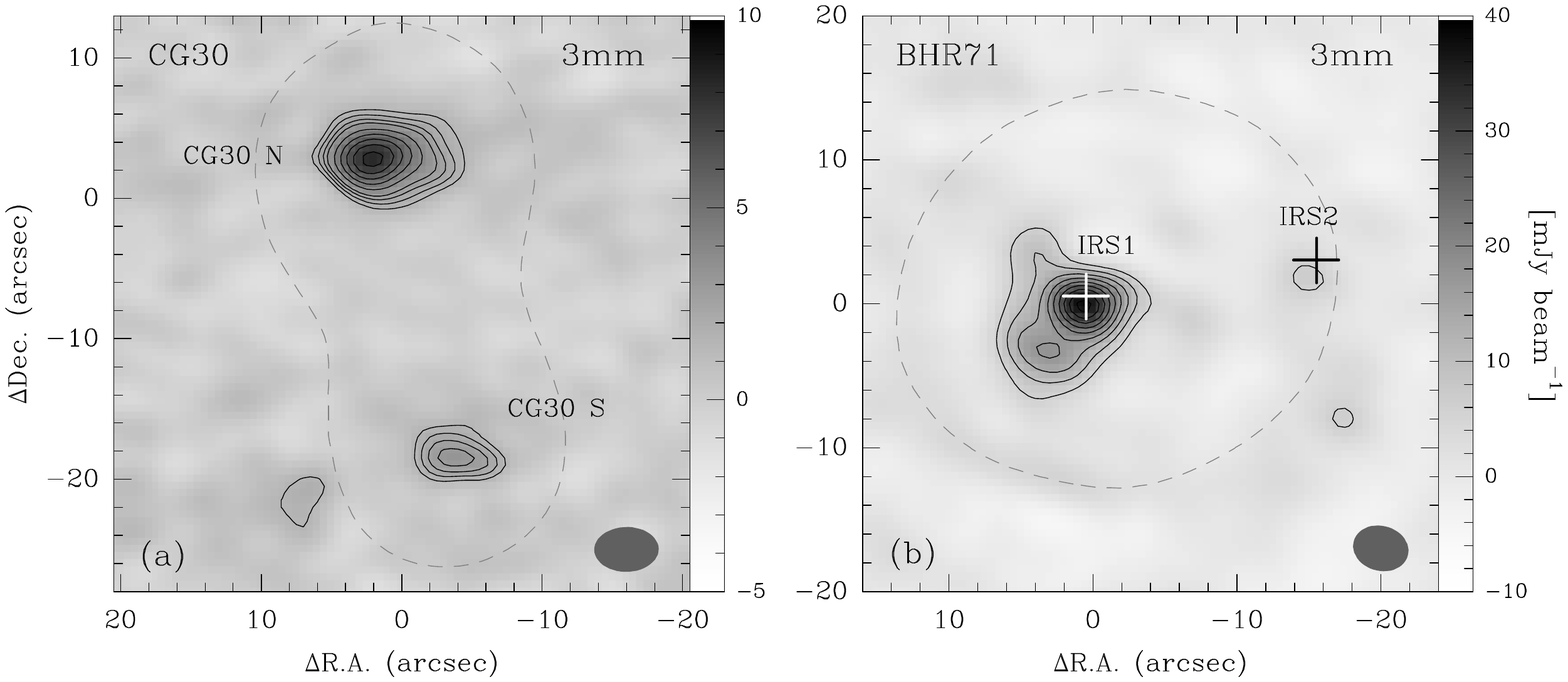}
\caption{(a) 3\,mm dust continuum image of CG\,30. Contours start
at $\sim$\,3\,$\sigma$ (1\,$\sigma$\,$\sim$\,0.5\,mJy) with steps
of $\sim$\,2\,$\sigma$. The grey dashed contour represents the
half-maximum level of the 850\,$\mu$m emission observed with SCUBA
(Henning et al. 2001). (b) 3\,mm dust continuum image of BHR\,71.
Contours start at $\sim$\,3\,$\sigma$
(1\,$\sigma$\,$\sim$\,2\,mJy) with steps of $\sim$\,2\,$\sigma$.
Crosses mark the positions of the $Spitzer$ MIR sources. The grey
dashed contour represents the half-maximum level of the 1.2\,mm
emission observed with SEST (B97). Synthesized ATCA beams are
shown as grey ovals in bottom right corners.\label{3mmc}}
\end{center}
\end{figure*}


\begin{figure*}
\begin{center}
\includegraphics[width=18cm,angle=0]{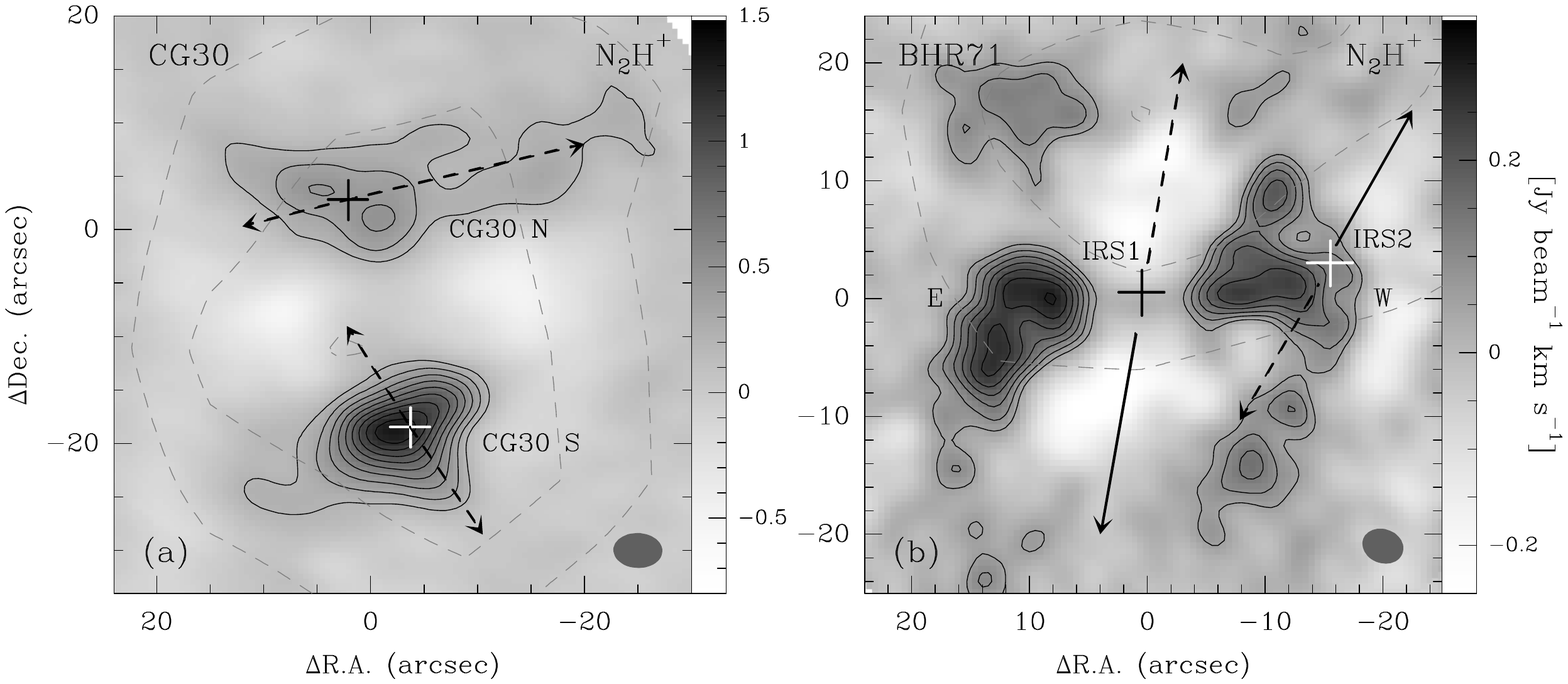}
\caption{(a) Image of the N$_2$H$^+$\,(1\,$-$\,0) intensity
integrated over the seven hyperfine components for CG\,30.
Contours start at $\sim$\,3\,$\sigma$
(1\,$\sigma$\,$\sim$\,60\,mJy) with steps of $\sim$\,2\,$\sigma$.
The arrows show the directions of protostellar jets (see
$\S$\,3.3). The grey dashed contours represent the 50\%, 75\%, and
99\% levels of the N$_2$H$^+$\,(1\,$-$\,0) emission observed with
Mopra single-dish telescope (P.~Barnes et al. in prep.). (b) The
same for BHR\,71 (1\,$\sigma$\,$\sim$\,20\,mJy). The solid and
dashed arrows show the directions of the blue-shifted and
red-shifted CO outflows (see $\S$\,3.3). The crosses in both
images represent the peaks of 3\,mm dust continuum emission.
Synthesized ATCA beams are shown as grey ovals.\label{n2hp}}
\end{center}
\end{figure*}

\begin{figure*}
\begin{center}
\includegraphics[width=18cm, angle=0]{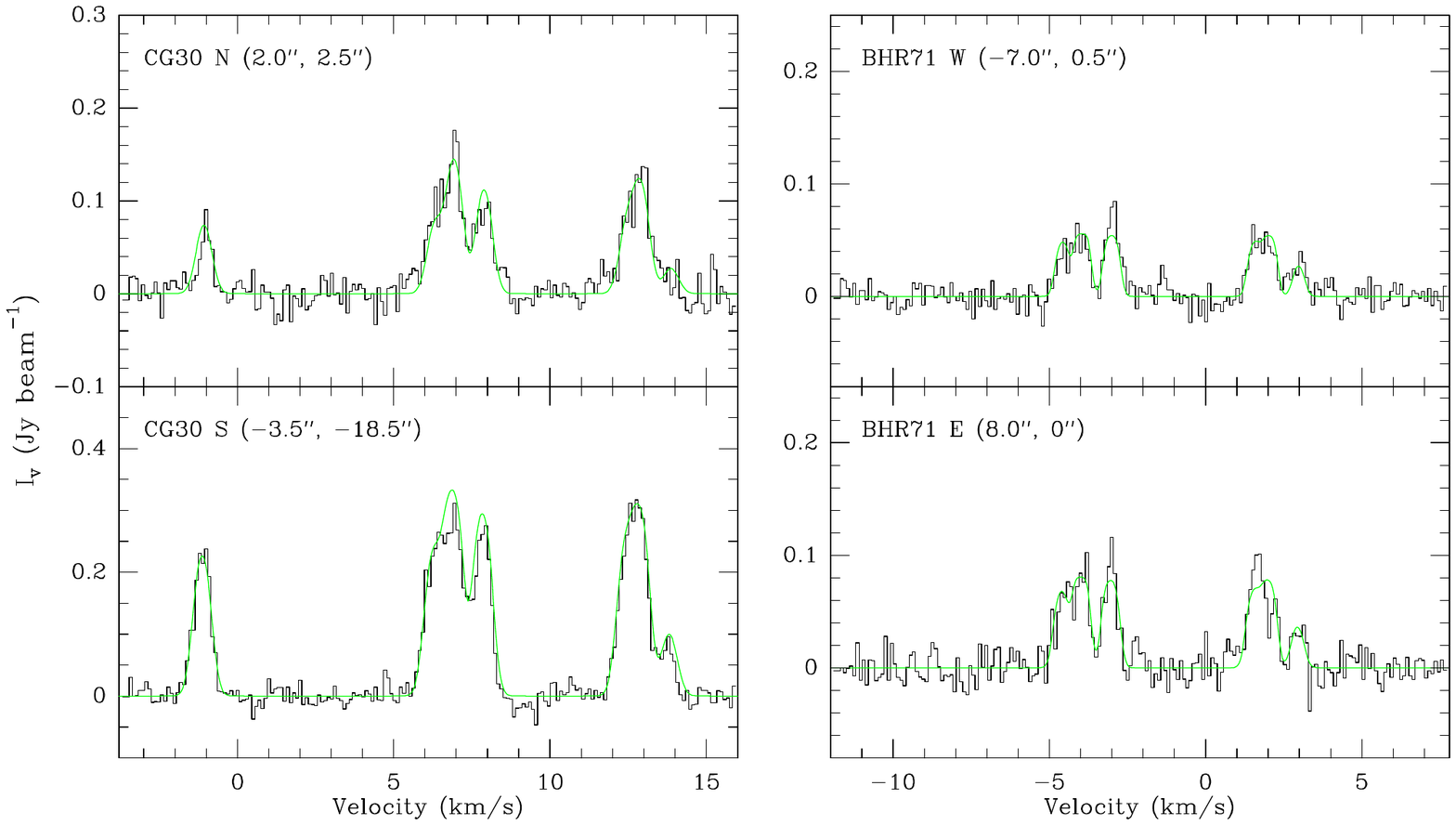}
\caption{N$_2$H$^+$ spectra at the peak positions of the two cores
in CG\,30 (left) and BHR\,71 (right). Thin dotted curves show the
results of hyperfine structure line fitting. Fit parameters are
given in Table~3. [{\it See the electronic edition of the Journal
for a color version of this figure.}] \label{spectra}}
\end{center}
\end{figure*}

\begin{figure*}
\begin{center}
\includegraphics[width=18cm, angle=0]{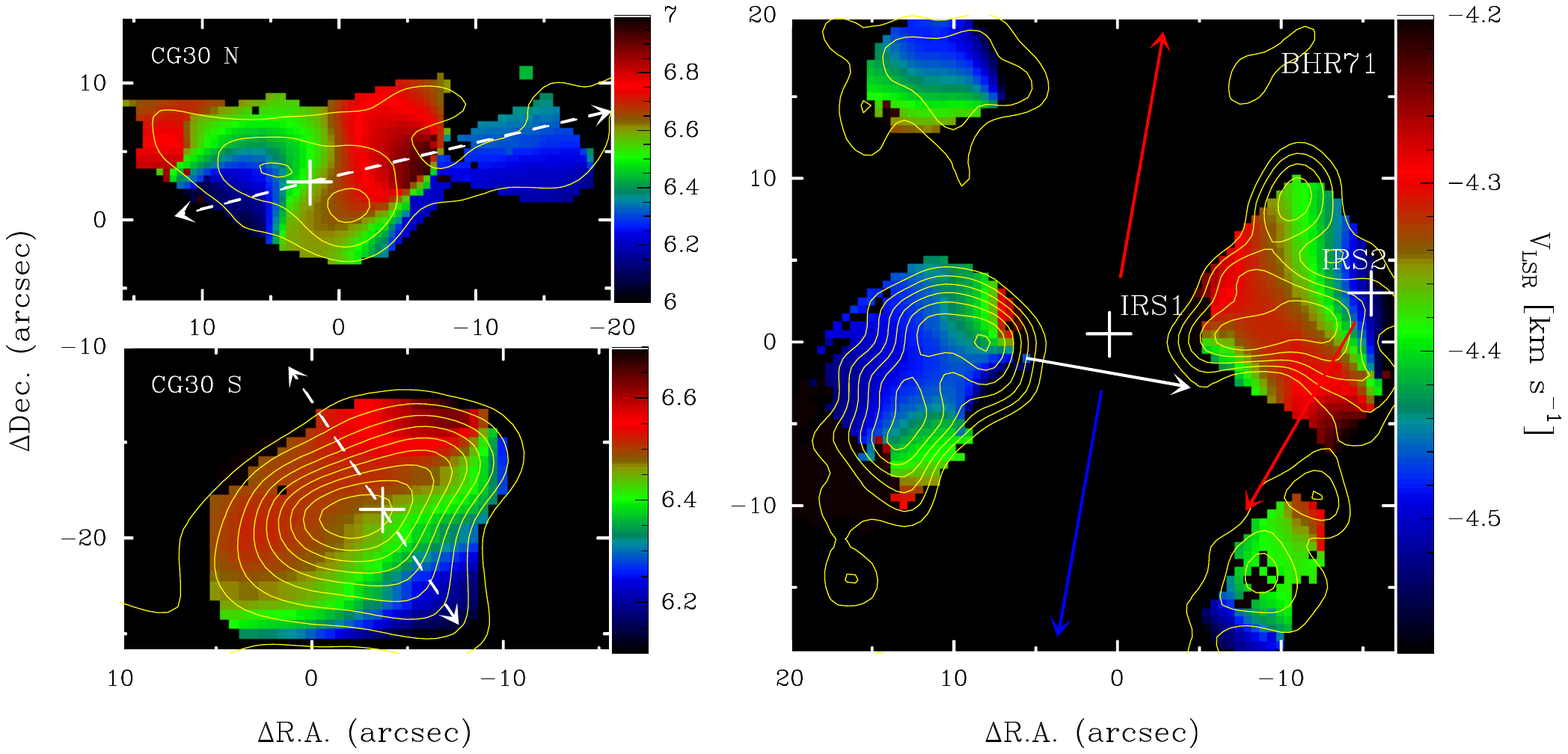}
\caption{N$_2$H$^+$ velocity field maps of CG\,30 (left) and
BHR\,71 (right). Contours are same as in Fig.\,2. The white arrows
in CG\,30N and CG\,30S show the directions of the protostellar
jets. The red and blue arrows in the BHR\,71 map show the
directions of CO outflows, while the white arrow shows the
direction of the gradient across the two main N$_2$H$^+$
clumps.\label{velocity}}
\end{center}
\end{figure*}


\begin{figure*}
\begin{center}
\includegraphics[width=18.0cm, angle=0]{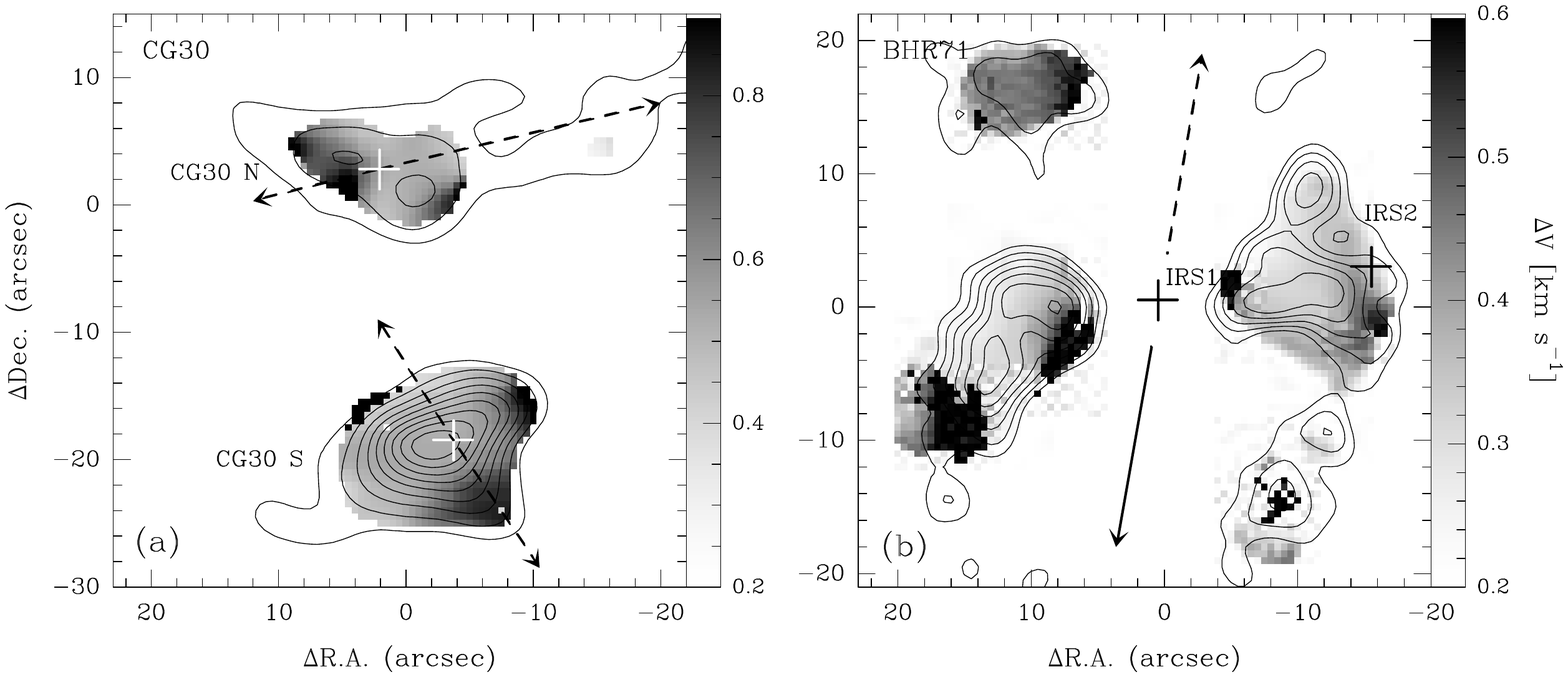}
\caption[]{Spatial distribution of N$_2$H$^+$ line widths in
CG\,30 (left) and BHR\,71 (right), as derived from the HFS line
fitting. Contours and symbols are the same as in Fig.\,2.
\label{linewidth1}}
\end{center}
\end{figure*}


\begin{figure*}
\begin{center}
\includegraphics[width=14.0cm, angle=0]{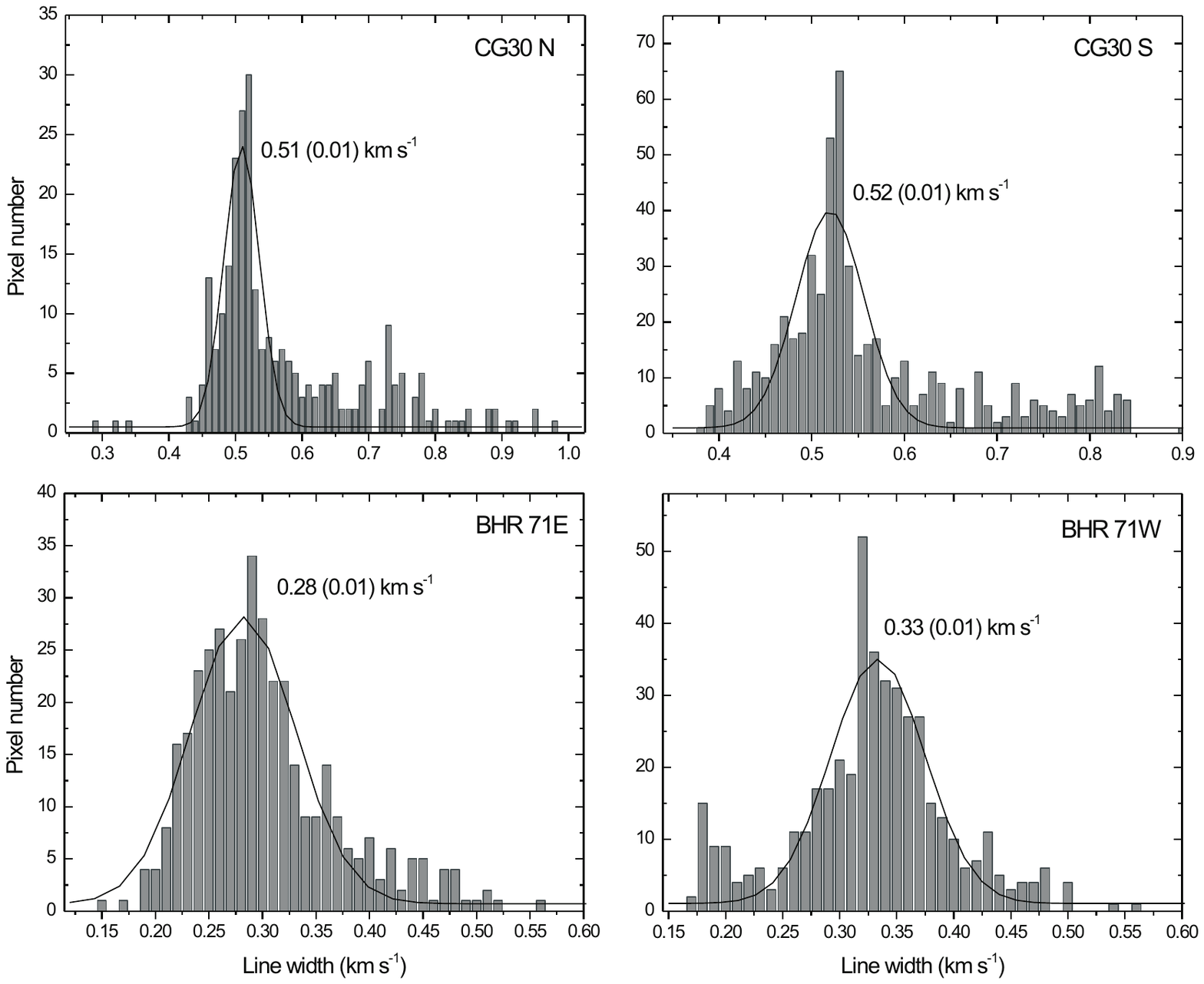}
\caption[]{Distribution of N$_2$H$^+$ line widths versus solid
angle areas for CG\,30 (top) and BHR\,71 (bottom). Black solid
curves and numbers show the results of Gaussian fitting to the
distributions.\label{linewidth2}}
\end{center}
\end{figure*}


\begin{figure*}
\begin{center}
\includegraphics[width=12cm,angle=0]{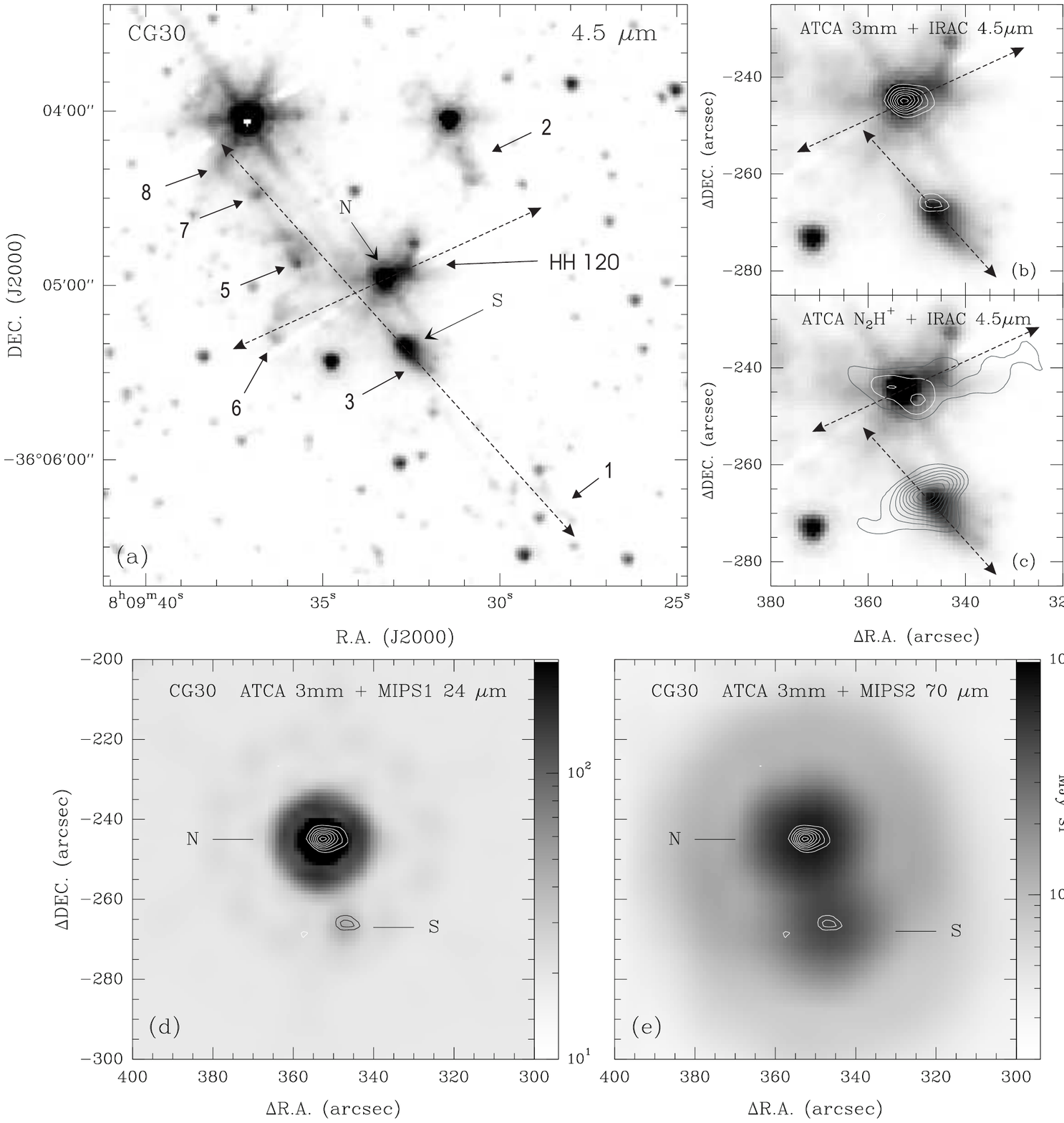}
\caption{\footnotesize $Spitzer$ images of CG\,30. (a) $Spitzer$ IRAC band 2
(4.5\,$\mu$m) image of CG\,30. Sources CG\,30N and CG\,30S are
labeled as ``N" and ``S", respectively. Dashed arrows show the
directions of the protostellar jets; (b) IRAC band 2 image
overlaid with the ATCA 3\,mm dust continuum contours (reference
position at R.A.=08:09:04.082, DEC=$-$36:00:53.53, J2000); (c)
Same, but overlaid with the ATCA N$_2$H$^+$ intensity contours;
(d) $Spitzer$ MIPS\,1 (24\,$\mu$m) image of CG\,30, overlaid with
the ATCA 3\,mm dust continuum contours; (e) $Spitzer$ MIPS\,2
(70\,$\mu$m) image of CG\,30, overlaid with the ATCA 3\,mm dust
continuum contours.\label{spitzer1}}
\end{center}
\end{figure*}


\begin{figure*}
\begin{center}
\includegraphics[width=14cm,angle=0]{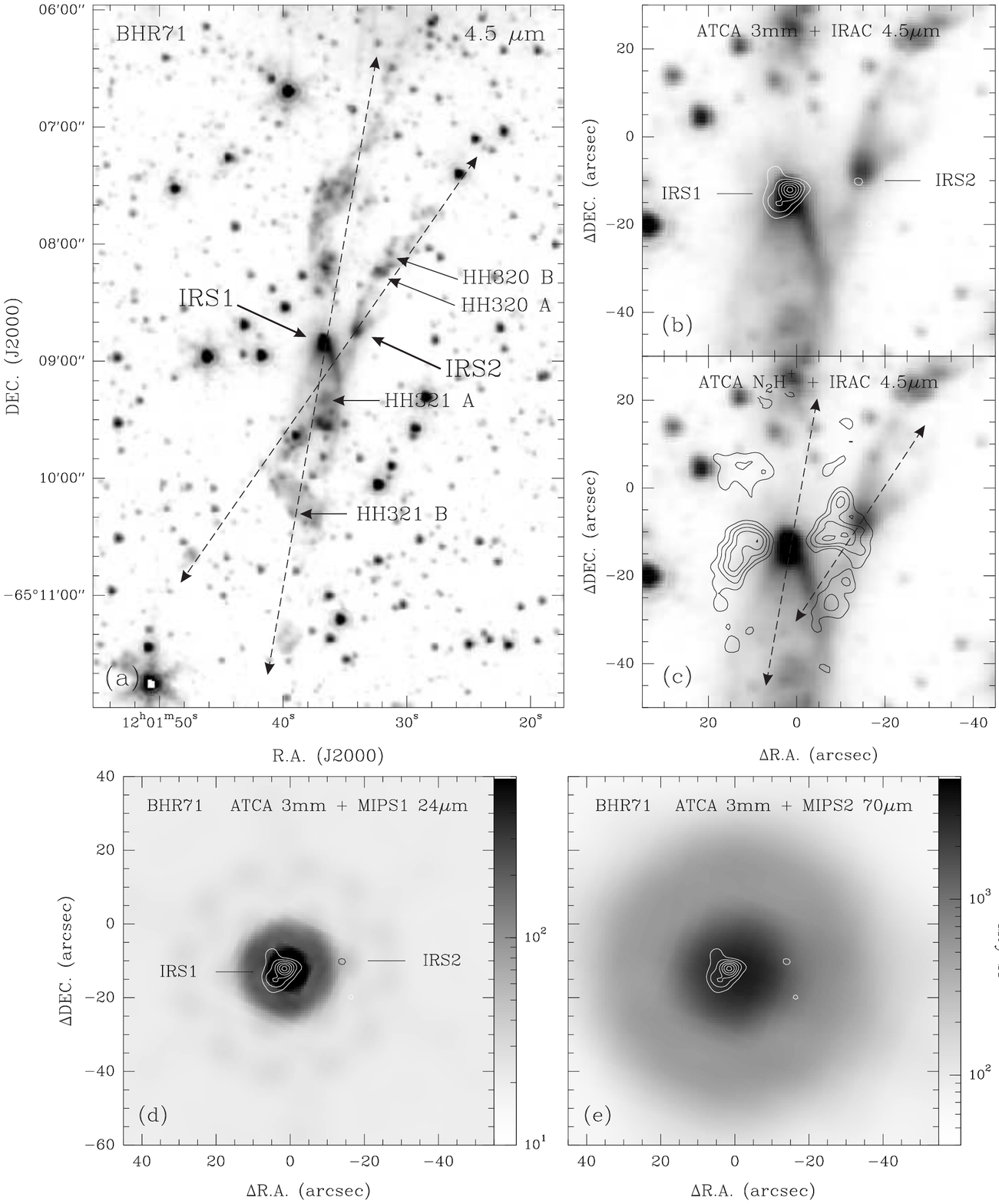}
\caption{The same as Fig.\,7, but for BHR\,71 (reference position
at R.A.=12:01:36.349, DEC=$-$65:08:37.50, J2000).\label{spitzer2}}
\end{center}
\end{figure*}


\begin{figure*}
\begin{center}
\includegraphics[width=12cm, angle=0]{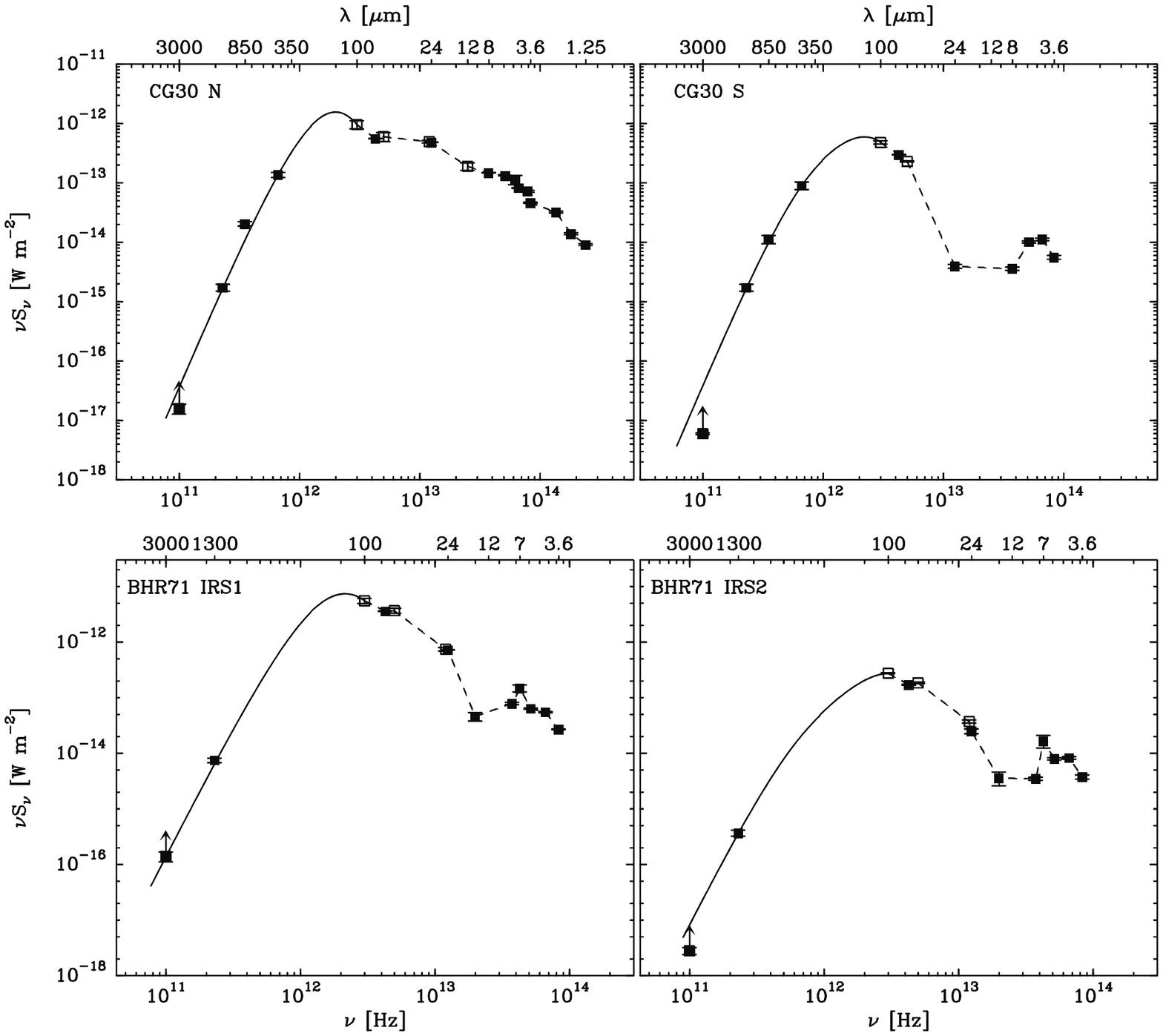}
\caption{\footnotesize Spectral energy distribution of CG\,30N (up left),
CG\,30S (up right), BHR\,71 IRS1 (bottom left), and BHR\,71 IRS2
(bottom right). Error bars (1\,$\sigma$) are indicated for all
data points, but are mostly smaller than the symbol sizes. Open
squares represent IRAS data points, where flux densities are
divided into two sub-cores with ratios assumed in $\S$\,4.1. While
most data points represent total fluxes, the 3\,mm fluxes were
measured from interferometric maps which resolved out the envelope
and thus represent lower limits only. Solid lines show the
best-fit for all points at $\lambda$\,$\geq$\,100\,$\mu$m using a
grey-body model. Dashed lines at $\lambda$\,$\leq$\,100\,$\mu$m
show the simple logarithmic interpolation used to derive the
luminosity. The fitting results are summarized in
Table~7.\label{sed}}
\end{center}
\end{figure*}

\begin{figure*}
\begin{center}
\includegraphics[width=14cm, angle=0]{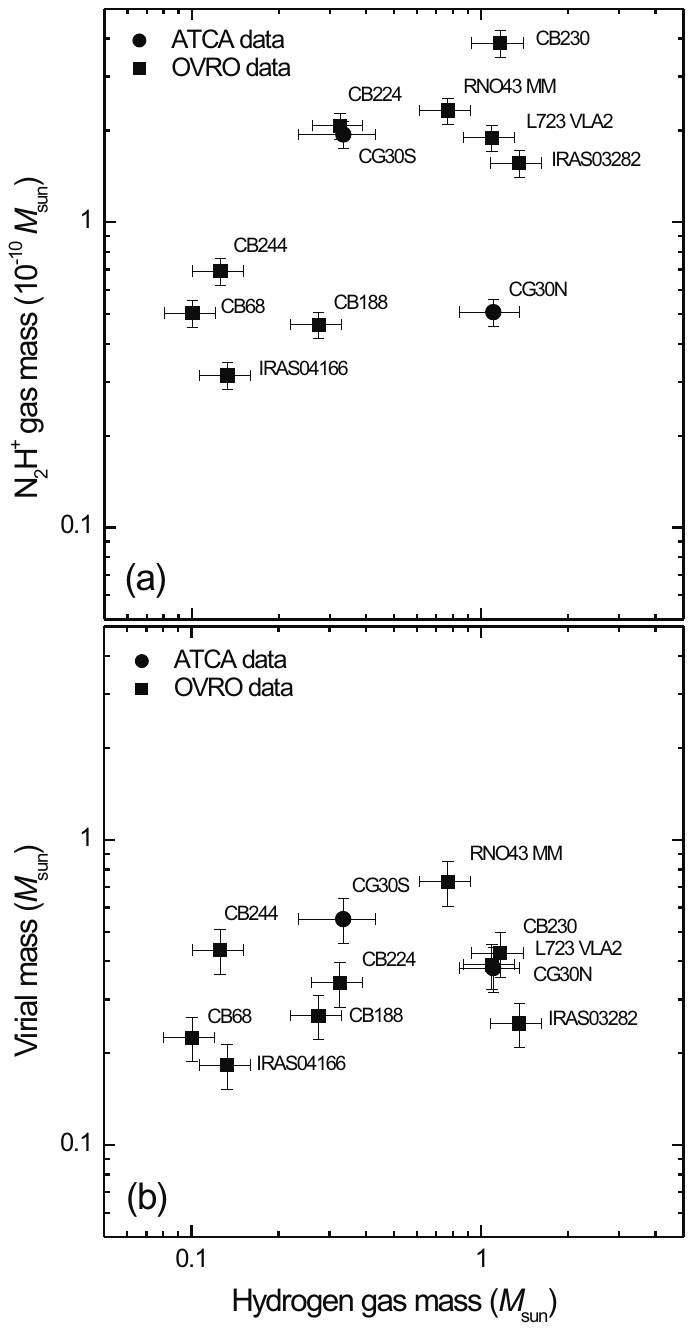}
\caption{\footnotesize (a) N$_2$H$^+$ gas mass (derived from N$_2$H$^+$ line
emission) versus hydrogen gas mass (derived from 3\,mm dust
continuum emission; the dust continuum data of OVRO sample are
taken from Launhardt et al. in prep.), and (b) virial mass
(derived from N$_2$H$^+$ line emission) versus hydrogen gas mass
for protostellar cores studied in Paper\,I and this
work.\label{last}}
\end{center}
\end{figure*}

\end{document}